\documentclass{emulateapj}
\usepackage{amsmath}
 
\newcommand{\Mstar}{\ifmmode M_{\ast} \else $M_{\ast}$}
\newcommand{\kms}{\ifmmode \mathrm{km~s^{-1}} \else km~s$^{-1}$\fi}
\newcommand{\smpy}{\ifmmode M_{\sun}~\mathrm{yr}^{-1} \else $M_{\sun}$~yr$^{-1}$\fi}
\newcommand{\msol}{\ifmmode M_{\sun} \else $M_{\sun}$\fi}
\newcommand{\smpc}{\ifmmode M_{\sun}~\mbox{pc}^{-2} \else $M_{\sun}$~pc$^{-2}$\fi}
\newcommand{\smkpc}{\ifmmode M_{\sun}~\mbox{yr}^{-1}~\mbox{kpc}^{-2} \else $M_{\sun}$~yr$^{-1}$~kpc$^{-2}$\fi}
\newcommand{\htwo}{\ifmmode \mbox{H}_{2} \else H$_{2}$\fi}
\newcommand{\ha}{\ifmmode \mbox{H}\alpha \else H$\alpha$\fi}
\newcommand{\sightwo}{\ifmmode \Sigma_{\textnormal{H}_{2}} \else $\Sigma_{\textnormal{H}_{2}}$\fi}
\newcommand{\sigmol}{\ifmmode \Sigma_{\textnormal{mol}} \else $\Sigma_{\textnormal{mol}}$\fi}
\newcommand{\sigsfr}{\ifmmode \Sigma_{\textnormal{SFR}} \else $\Sigma_{\textnormal{SFR}}$\fi}
\newcommand{\siggas}{\ifmmode \Sigma_{\textnormal{gas}} \else $\Sigma_{\textnormal{gas}}$\fi}
\newcommand{\sighi}{\ifmmode \Sigma_\text{H\textsc{i}} \else $\Sigma_\text{H\textsc{i}}$\fi}
\newcommand{\sigdust}{\ifmmode \Sigma_\text{dust} \else $\Sigma_\text{dust}$\fi}
\newcommand{\sigstar}{\ifmmode \Sigma_{*} \else $\Sigma_{*}$\fi}
\newcommand{\sigdotstar}{\ifmmode \dot{\Sigma}_{*} \else $\dot{\Sigma}_{*}$\fi}
\newcommand{\cplus}{\ifmmode \mbox{C}^{+} \else C$^{+}$\fi}
\newcommand{\hi}{{\sc H\,i}}
\newcommand{\hii}{{\sc H\,ii}}

\newcommand{\he}{\hbox{{He}}}
\newcommand{\Nhi}{\ifmmode N_\textnormal{H\textsc{i}} \else $N_\textnormal{H\textsc{i}}$\fi}
\newcommand{\Nmol}{\ifmmode N_\text{mol} \else $N_\text{mol}$\fi}
\newcommand{\Nhtwo}{\ifmmode N_{\textnormal{H}_{2}} \else $N_{\textnormal{H}_{2}}$\fi}
\newcommand{\Mhi}{\ifmmode M_\text{H\textsc{i}} \else $M_\text{H\textsc{i}}$\fi}
\newcommand{\Mhtwo}{\ifmmode M_{\textnormal{H}_{2}} \else $M_{\textnormal{H}_{2}}$\fi}
\newcommand{\Mmol}{\ifmmode M_{\textnormal{mol}} \else $M_{\textnormal{mol}}$\fi}
\newcommand{\dGDR}{\ifmmode \delta_{\textnormal{GDR}} \else $\delta_{\textnormal{GDR}}$\fi}
\newcommand{\co}{$^{12}$CO}
\newcommand{\taudep}{\ifmmode \tau_{\textnormal{dep}} \else $\tau_{\textnormal{dep}}$\fi}
\newcommand{\taumoldep}{\ifmmode \tau^{\textnormal{mol}}_{\textnormal{dep}} \else $\tau^{\textnormal{mol}}_{\textnormal{dep}}$\fi}

\newcommand{\myemail}{kjameson@astro.umd.edu}

\slugcomment{Accepted for publication in ApJ 4/25/2016}

\shorttitle{Gas and Star Formation in the Magellanic Clouds}
\shortauthors{Jameson et al.}

\begin{document}


\title{The Relationship Between Molecular Gas, \hi, and Star Formation in the Low-Mass, Low-Metallicity Magellanic Clouds}

\author{Katherine E. Jameson \altaffilmark{1}, Alberto D. Bolatto\altaffilmark{1}, Adam K. Leroy \altaffilmark{2}, Margaret Meixner \altaffilmark{3}, Julia Roman-Duval \altaffilmark{3}, Karl Gordon \altaffilmark{3}, Annie Hughes \altaffilmark{4}, Frank P. Israel \altaffilmark{5}, Monica Rubio\altaffilmark{6},  Remy Indebetouw \altaffilmark{7,8}, Suzanne C. Madden \altaffilmark{9}, George Sonneborn \altaffilmark{10}, Caroline Bot \altaffilmark{11}, Maud Galametz \altaffilmark{12}, Sacha Hony \altaffilmark{13}, Diane Cormier \altaffilmark{13}, Eric W. Pellegrini \altaffilmark{14}}

\email{\myemail}

\altaffiltext{1}{Astronomy Department and Laboratory for Millimeter-wave Astronomy, University of Maryland, College Park, MD 20742}
\altaffiltext{2}{Department of Astronomy, The Ohio State University, 4051 McPherson Laboratory, 140 West 18th Avenue, Columbus, OH 43210, USA}
\altaffiltext{3}{Space Telescope Science Institute, 3700 San Martin Dr., Baltimore, MD 21218, USA}
\altaffiltext{4}{Max-Planck-Institut f\"{u}r Astronomie, K\"{o}nigstuhl 17 D-69117 Heidelberg, Germany}
\altaffiltext{5}{Sterrewacht Leiden, Leiden University, PO Box 9513, 2300 RA Leiden, The Netherlands} 
\altaffiltext{6}{Departamento de Astronom'a, Universidad de Chile, Casilla 36-D, Chile}
\altaffiltext{7}{Department of Astronomy, University of Virginia, PO Box 400325, Charlottesville, VA 22904, USA}
\altaffiltext{8}{National Radio Astronomy Observatory, 520 Edgemont Road, Charlottesville, VA 22903, USA}
\altaffiltext{9}{Laboratoire AIM, CEA, UniversitŽ Paris VII, IRFU/Service dÕAstrophysique, Bat. 709, 91191 Gif-sur-Yvette, France}
\altaffiltext{10}{NASAÕs Goddard Space Flight Center, Observational Cosmology Laboratory, Code 665, Greenbelt, MD 20771}
\altaffiltext{11}{Observatoire astronomique de Strasbourg, UniversitŽ de Strasbourg, CNRS, UMR 7550, 11 rue de l'UniversitŽ, 67000 Strasbourg, France}
\altaffiltext{12}{European Southern Observatory, Karl-Schwarzschild-Str. 2, D-85748 Garching-bei-MŸnchen, Germany}
\altaffiltext{13}{Institut fŸr Theoretische Astrophysik, Zentrum fŸr Astronomie, UniversitŠt Heidelberg, Albert-Ueberle-Str. 2, D-69120 Heidelberg, Germany}
\altaffiltext{14}{Department of Physics and Astronomy, University of Toledo, 2801 West Bancroft Street, Toledo, OH 43606, USA}


\begin{abstract}

The Magellanic Clouds provide the only laboratory to study the effect of metallicity and galaxy mass on molecular gas and star formation at high ($\sim{20}$ pc) resolution. We use the dust emission from HERITAGE $Herschel$ data to map the molecular gas in the Magellanic Clouds, avoiding the known biases of CO emission as a tracer of \htwo. Using our dust-based molecular gas estimates, we find molecular gas depletion times (\taumoldep) of $\sim{0.4}$ Gyr in the LMC and $\sim{0.6}$ SMC at 1 kpc scales. These depletion times fall within the range found for normal disk galaxies, but are shorter than the average value, which could be due to recent bursts in star formation. We find no evidence for a strong intrinsic dependence of the molecular gas depletion time on metallicity. We study the relationship between gas and star formation rate across a range in size scales from 20 pc to $\geq 1$ kpc, including how the scatter in \taumoldep\ changes with size scale, and discuss the physical mechanisms driving the relationships. We compare the metallicity-dependent star formation models of \citet{ost10} and \citet{kru13} to our observations and find that they both predict the trend in the data, suggesting that the inclusion of a diffuse neutral medium is important at lower metallicity. 

\end{abstract}



\keywords{galaxies: dwarf -- galaxies: evolution -- ISM: clouds -- Magellanic Clouds}

\section{Introduction}

Star formation plays a critical role in shaping how galaxies form and evolve. Understanding the molecular gas content of low-mass, low-metallicity galaxies and its relationship to the star formation rate is necessary to understand how the gas mass fractions evolve with redshift \citep[e.g.,][]{tac10, gen12} and how the star formation efficiency depends on galaxy mass and metallicity \citep[e.g.,][]{bol11, sai11, kru11}. Both are critical to understanding the ``galaxy mass function''  and drivers of the star formation history of the universe. 

Our current knowledge of the extragalactic relationship between gas and star formation comes from studies of mostly high-metallicity, high-mass nearby galaxies that use \co\ to trace the molecular gas. The original work to quantitatively compare the star formation rate to the gas density by \citet{sch59} found a power law relationship, generally referred to as the ``star formation law.'' More recent studies of the extragalactic star formation law follow the work of \citet{ken89,ken98}, which used primarily disk-averaged measurements of the surface density of total gas ($\siggas=\sightwo+\sighi$) and star formation rate (\sigsfr). They found that the relationship between \siggas\ and \sigsfr\ follows a power law distribution ($\sigsfr\propto\siggas$$^{1+p}$). Studies at higher resolution found that the general power law trend continued, but only within the molecular-dominated regimes and that \sigmol\ and \sigsfr\ follow an approximately linear power law relation \citep{big08,sch11,big11,rah12}. At lower gas surface densities, where \hi\ dominates the total gas budget, they see a steep fall off in the relationship. Resolved galaxy studies show that while the total gas continues to be correlated with the star formation rate within galaxies, the molecular gas correlates best with the star formation rate. 

Due to the nearly linear power law slope of the relationship between \sigmol\ and \sigsfr, a convenient way to quantify the relationship is the molecular gas depletion time: $\taumoldep=\sigmol/\sigsfr$. The depletion time can be thought of as the amount of time it would take to deplete the current reservoir of molecular gas given the current star formation rate. Most of the resolved data for samples of galaxies achieve resolutions of several hundred parsecs to $\sim{1}$ kpc \citep{big08,ler08,big11,rah12,ler13a} and all find similar values for the average molecular gas depletion time of $\taumoldep \sim{2}$ Gyr. The weak dependence of \taumoldep\ on the galactic properties and environment \citep{ler13a} suggests that star formation is a local process based on the conditions within giant molecular clouds (GMCs). 

The conclusions from studies of mostly high-mass, high-metallicity disk galaxies may not extend to lower metallicity star-forming dwarf galaxies where the ISM is dominated by atomic gas. The lack of metals produce different physical conditions that potentially affect the molecular gas fraction and how star formation proceeds within the galaxy. For example, the galaxies will have lower dust-to-gas ratios, which results in lower extinctions and higher photodissociation rates. The Large and Small Magellanic Clouds (LMC, SMC) provide ideal laboratories to study the physics of star formation at low mass, $M_{\textnormal{*,LMC}}=2\times10^{9}$ \msol\ and $M_{\textnormal{*,SMC}}=3\times10^{8}$ \msol\ \citep{ski12}, and low metallicity, $Z_{\textnormal{LMC}}\sim{1/2}~Z_{\sun}$ \citep{rus92} and $Z_{\textnormal{SMC}}\sim{1/5}~Z_{\sun}$ \citep{duf84,kur99,pag03}, due to their proximity and our ability to achieve high spatial resolution ($\sim$ 10 pc).

Tracing the molecular gas at low-metallicity is difficult because CO, the most common tracer of \htwo, emits weakly and is often undetected. The Magellanic Clouds have been studied extensively in \co\ with the earliest surveys completed using the Columbia 1.2m \citep{coh88,rub91}. The early survey of both Clouds completed by \citet{isr93} using the Swedish-ESO Submillimetre Telescope (SEST) showed the CO emission to be under-luminous compared to the Milky Way by a factor of $\sim{3}$ in the LMC and $\sim{10}$ in the SMC. Since then, many large-scale surveys have been completed for the LMC \citep{fuk08,won11} and SMC \citep{rub93,miz01,mul10}. The \htwo\ gas is expected to be more prevalent than CO at low metallicity due to increased ability of \htwo\ to self-shield against dissociating UV photons compared to CO. Both observations and modeling suggest that $\sim{}30\% - 50\%$ of the H$_{2}$ in the Solar Neighborhood resides in a ``CO-faint'' phase \citep[e.g.,][]{gre05, wol10, pla11}, similar to the estimated fraction of in the LMC \citep[e.g.,][]{rom10, ler11}. Studies of the SMC find this phase to encompass $80\% - 90\%$ of all the H$_{2}$ \citep{isr97, pak98, ler07, ler11, bol11}, likely dominating the molecular reservoir available to star formation. 

Using dust emission to estimate the molecular gas in low metallicity systems avoids the biases of CO and can trace ``CO-faint'' molecular gas. This method of tracing the molecular gas using dust emission in the Magellanic Clouds was first applied by \citet{isr97} using $IRAS$ data, and later by \citet{ler07, ler09} in the SMC and \citet{ber08} in the LMC using $Spitzer$ data. \citet{bol11} further refined the methodology and created a map of \htwo\ in the SMC using dust continuum emission from $Spitzer$ and studied the spatial correlation between the atomic gas, molecular gas, and star formation rate. When using the dust-based molecular gas estimate, they found that \taumoldep\ is consistent with the values seen in more massive disk galaxies. Combining the dust-based molecular gas estimate with the atomic gas traced by \hi\ showed that the analytic star formation models of \citet{kru09} and \citet{ost10} predicted the trend in the data.

In this work we produce an estimate of the molecular gas using dust emission traced by $Herschel$ in the LMC and SMC. While the SMC is lower metallicity, the geometry is poorly constrained and it shows clear signs of disturbance from  interaction with the LMC and the Milky Way, which makes it problematic for comparisons against models created for galactic disks. We adopt a higher inclination angle for the SMC than was used in \citet{bol11} to explore how that affects the results. The LMC is nearly face-on  with a well-constrained inclination angle and has a clear disk morphology, which minimizes the uncertainty in the analysis. 

We compare the new dust-based molecular gas estimates and atomic gas to the star formation rate in both galaxies and compare to the existing studies of large disk galaxies. In Sections \ref{section:observations} and \ref{section:methodology} we outline the observations and how we convert them to physical quantities. Section \ref{section:results} presents the main results of this study, focusing on the relationship between molecular gas and star formation and the effect of scale. We discuss the implications of the results and compare the observations to star formation model predictions in Section \ref{section:discussion}. Finally, we summarize the conclusions from this study of the LMC and SMC in Section \ref{section:conclusions}.

\section{Observations}
\label{section:observations}

\subsection{Herschel Data}
\label{subsection:Herschel_data}

The far-infrared images come from the HERschel Inventory of The Agents of Galaxy Evolution in the Magellanic Clouds key project \citep[HERITAGE;][]{mei13}. HERITAGE mapped both the Large Magellanic Cloud (LMC) and Small Magellanic Cloud (SMC) at 100, 160, 250, 350, and 500 \micron\ with the Spectral and Photometric Imaging Receiver (SPIRE) and Photodetector Array Camera and Spectrometer (PACS) instruments. Information on the details of the data calibration, reduction, and uncertainty can be found in \citet{mei13}. 

For this work, we apply further background subtraction. First, we remove the foreground Milky Way cirrus emission. Following \citet{gor14} (based on \citealt{bot04}), we estimate the foreground cirrus emission by using the relationship between IR dust emission and \hi\ from \citet{des90} and scaling the integrated \hi\ intensity map over the velocities of the Milky Way emission in the direction of the LMC by applying the conversion factors 1.073, 1.848, 1.202, 0.620 (MJy sr$^{-1}/10^{2-}$ cm${-2}$) for 100 \micron, 160 \micron, 250 \micron, and 350 \micron, respectively. The median estimated cirrus emission was 5.7, 9.9, 6.4, and 3.3 MJy sr$^{-1}$ for the 100 \micron, 160 \micron, 250 \micron, and 350 \micron\ images. 

Second, we set the images to comparable zero-points: the outskirts of the PACS images were set to the COBE and IRAS data emission levels while the outskirts of the SPIRE images were set to zero due to the lack of similar large-scale coverage at the longer wavelengths. After subtracting the cirrus emission, we chose 6 regions in the outskirts of the LMC with no emission in the $Herschel$ or \hi\ images, fit a plane to the median values of the regions and subtract the plane. The cirrus-subtraction and background-subtraction had primarily minor effects on the images, with the final image values being lower by 7\%, 9\%, 2\%, and 2\% on average for the 100 \micron, 160 \micron, 250 \micron, and 350 \micron\ images in regions with $S/N>3$. 

\subsection{\hi\ Data}
\label{subsection:HI_data}

The neutral atomic gas data come from 21 cm line observations of \hi. We use the LMC \hi\ map from \citet{kim03} and the SMC \hi\ map from \citet{sta99}, both combined Australian Telescope Compact Array (ATCA) and Parkes 64m radio telescope data. The interferometric ATCA data set the map resolution at $1\arcmin$ ($r\sim20$ pc in the SMC and $r\sim15$ pc in the LMC), but the data are sensitive to all size scales due to the combination of interferometric and single-dish data. 

The observed brightness temperature of the 21 cm line emission is converted to \hi\ column density (\Nhi) assuming optically thin emission using 
\[ \Nhi = 1.823\times{10^{18}}\frac{\mbox{ cm}^{-2}}{\mbox{K km s}^{-1}} \int T_{B}(v)~\mathrm{d}v\,. \]
We find RMS column densities of $8.0\times10^{19}$ cm$^{-2}$ in the LMC map and $5.0\times10^{19}$ cm$^{-2}$ in the SMC map. We convert column density to surface mass density (\sighi) using
\[ \sighi =  1.4\cos{i}\left(8.0\times10^{-21}\frac{\msol \mbox{ pc}^{-1}}{\mbox{cm}^{-2}}\right)\Nhi \,,\]
where the factor of 1.4 accounts for \he\ and $i$ is the inclination angle.

While the assumption of optically thin \hi\ emission is likely appropriate throughout much of the galaxies, there are regions with optically thick emission, which would cause \Nhi\ to be underestimated. While a statistical correction for \hi\ optical depth in the SMC exists \citep{sta99}, none exists for the LMC. Additionally, nearby surveys of \hi\ (i.e., THINGS; \citealt{wal08}) make no optical depth corrections. We chose not to make any optical depth corrections to the \hi\ maps as the statistical corrections in the SMC are generally small (increases the total \hi\ mass by $10\%$; \citealt{sta99}) and an accurate optical depth correction would require assuming a spin temperature. 
\subsection{CO Data}
\label{subsection:CO_data}

We use integrated \co\ $(1-0)$ intensity maps from the 4m NANTEN radio telescope (half power beam width of 2.6\arcmin\ at 115 GHz) for the LMC \citep{fuk08} and SMC \citep{miz01}. The LMC and SMC velocity integrated maps have typical $3\sigma$ noise of $\sim{1.2}$ K km s$^{-1}$ and $\sim{0.45}$ K km s$^{-1}$, respectively. For the LMC, there is also the higher resolution and sensitivity MAGellanic Mopra Assessment (MAGMA) Survey, which used the 22m Mopra telescope of the Australia Telescope National Facility to follow-up the NANTEN survey with 40\arcsec\ angular resolution and $1\sigma$ sensitivity of 0.2 K km s$^{-1}$ \citep{won11}. However, the MAGMA survey is not complete as they only mapped regions with detected CO in the NANTEN map. Because the CO maps are only used to identify molecular regions and do not affect the final resolution of our molecular gas maps, we use the higher coverage NANTEN maps in our molecular gas mapping process and then, in the LMC, we compare the final dust-based molecular gas maps to the higher resolution MAGMA data.  

\subsection{\ha\ and $Spitzer$ 24 \micron\ Data}
\label{subsection:Ha_data}

We combine images of \ha\ and 24 \micron\ dust emission to trace recent star formation. For the LMC we use the calibrated, continuum-subtracted H$\alpha$ map from the Southern H$\alpha$ Sky Survey Atlas \citep[SHASSA;][]{gau01} at 0.8\arcmin\ resolution. We correct the \ha\ maps for the line-of-sight Milky Way extinction using $A_{V}(\text{LMC}) = 0.2$ mag and $A_{V}(\text{SMC}) = 0.1$ mag \citep{sch11b}. We found background emission outside the LMC, on the order of 10\% of the total flux observed in the main part of the galaxy, likely from the diffuse Milky Way H$\alpha$ emission. We apply additional background subtraction by removing a polynomial fit to the regions outside of the galaxy. In the SMC, we use the continuum-subtracted H$\alpha$ map from the Magellanic Cloud Emission Line Survey \citep[MCELS;][]{smi99} at 2.3\arcsec\ resolution. For both the SMC and LMC we use the Multiband Imaging Photometer (MIPS) 24 \micron\ map from the $Spitzer$ Survey ``Surveying the Agents of Galaxy Evolution'' \citep[SAGE;][]{mei06,gor11}. 

\subsection{Distances and Inclination Angles}
\label{subsection:inclination}

To convert observational measurements to surface mass density ($\Sigma$), we need both the distance to the galaxy and inclination angle ($i$). For the LMC, we use an inclination angle of $i=35\degr$, which is the approximate intermediate value of the three fits to stellar proper motions and line-of-sight velocity measurements in \citet{van14}, which range from $i=26.2\degr\pm5.9\degr$ to $i=39.6\degr\pm4.5\degr$, and is consistent with their previous work that found $i=34.7\degr\pm6.2\degr$ \citep{van01}. We assume that the inclination of stellar disk is comparable to the gas disk given the disk-like morphology of the LMC. While \citet{kim98} fit an inclination angle to the \hi\ kinematics, they found it was unreliable and much higher than the morphological fit ($i=22\degr\pm6\degr$). Ultimately, \citet{kim98} adopt the inclination angle found from the stellar dynamics. 

The inclination of the SMC is poorly constrained due to its irregular morphology. Recent work by \citet{sco16} shows that assuming a disk with an inclination angle inaccurately represents the detailed morphology of the SMC. However, comparing the SMC to the LMC and studies of other galaxies requires knowing the mass surface densities and adopting the simple model of an inclined disk. \citet{bol11} adopted $i=40\degr\pm20\degr$ based on the analysis of the \hi\ rotation curve by \citet{sta04}. The recent estimate of the SMC inclination based on three dimensional structure traced by cepheid variable stars finds $i=74\degr\pm9\degr$ \citep{has12}, which is consistent with the previous studies using cepheids \citep{cal86,gro00}. While cepheids, as old stars, may not trace the gaseous disk, a new analysis of the \hi\ rotation also indicates a higher possible inclination of $i\approx60-70\degr$ (private communication, P. Teuben). A higher inclination angle scales the surface mass densities to lower values. We adopt $i=70\degr$ for the inclination of the SMC and compare to the previous results in \citet{bol11} to determine how the higher inclination angle affects the results. 

\section{Methodology}
\label{section:methodology}

\begin{figure}[t]
\hspace{-0.55cm}
\includegraphics[scale=0.75]{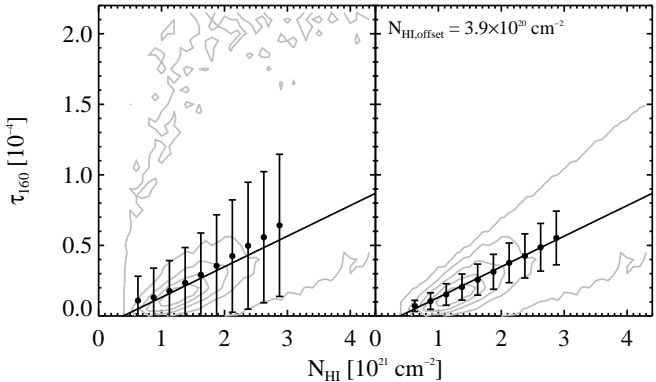}
\caption{Both plots show the relationship between \Nhi\ and $\tau_{160}$ (from the BEMBB dust modeling) in the LMC with the right plot showing the best representation of the relationship between \Nhi\ and $\tau_{160}$ as the lines of sight with molecular gas have been removed. The contours levels correspond to the full extent of the distribution, 20\%, 40\%, 60\%, and 80\% of the maximum density of points. The black points show the medians in $2.5\times10^{20}$ cm$^{-2}$ \Nhi\ bins with error bars showing $1\sigma$ of the distribution of measurements within the bin. The plots show the two-stage iteration used to fit the offset in the distribution: the $left$ plot has points near bright CO emission masked and the $right$ plot has masked points near bright CO and points with estimated $\Nmol>0.5\Nhi$ based on the first iteration. The typical error on $\tau_{160}$ is $\sim{1\times10^{-5}}$ in regions with predominately \hi\ gas, which is similar to the $1\sigma$ spread in the distribution in the bins. This suggest that the correlation between \Nhi\ and $\tau_{160}$ is intrinsically very tight and approximately linear, showing that the dust is a good tracer of the gas. We find the offset in \Nhi\ from the fit to the medians in the second iteration. The \Nhi\ offset is $\sim5\times10^{20}$ cm$^{-2}$ throughout most of the LMC (see Appendix \ref{appendix:hi_offset} for further details). 
\label{fig:lmc_Nhi_tau}}
\end{figure}

\subsection{Estimating Molecular Gas from Infrared Dust Emission}
\label{subsection:h2_method}

We combine the dust emission with a self-consistently estimated gas-to-dust ratio to estimate the total amount of gas. By removing the atomic gas, we are left with an estimate of the amount of molecular gas. The benefit of this method, particularly at low metallicity, is its ability to trace \htwo\ where CO has photo-dissociated. This method is based on the previous work by \citet{isr97} and \citet{ler11} in both Magellanic Clouds, \citet{dam01} in the Milky Way, \citet{ber08} in the LMC, and \citet{ler07,ler09} and \citet{bol11} in the SMC, all of which have demonstrated that dust is a reliable tracer of the molecular gas. In Figure \ref{fig:lmc_Nhi_tau}, we show that the optical depth of the dust correlates well with \Nhi, which represents the majority of the gas, and the $1\sigma$ scatter in the distribution is comparable to the uncertainty of $\tau_{160}$ of $\sim{1\times{10}^{-5}}$, suggesting there is intrinsically a tight relationship. The variation in the relationship between \Nhi\ and $\tau_{160}$ that is observed in nearby clouds (see below) could contribute to the observed scatter. The dust is also well correlated with the molecular gas traced by \co, which is shown for the SMC in Figure 4 in \citet{lee15a} using the HERITAGE and MAGMA data. We summarize the specific steps in our methodology, which closely follow the methodology by \citet{ler09} for the SMC, but with improvements allowed by the increased IR coverage and resolution from $Herschel$.

Following \citet{ler09} and \citet{bol11}, we model the dust emission in order to get the optical depth of the dust emission at 160 \micron\ ($\tau_{160}$). We use the results from two different dust emission fitting techniques for the LMC, one presented in this paper and another from \citet{gor14}, both based on the assumption of modified blackbody emission, $S_{\nu{}}\propto{\nu{^{\beta}}B_{\nu{}}}\left(T_{d}\right)$. We describe the fitting techniques in more detail in Appendix \ref{appendix:modeling_dust}. For the SMC, we only produce one molecular gas map using the modeling results from \citet{gor14} since \citet{bol11} produced a molecular gas map using a fixed $\beta$ simple modified blackbody model and a similar methodology. The \citet{gor14} dust modeling may produce a more accurate measure of $T_{d}$ since it allows $\beta$ to vary while reducing the amount of degeneracy between $T_{d}$ and $\beta$ \citep{dup03, she09} by accounting for the correlated errors between the $Herschel$ bands. 

While the dust temperature along the line of sight throughout the Magellanic Clouds likely has a distribution of temperatures \citep{ber08,gal11,gal13}, the assumption of a single dust temperature on the small spatial scales we cover ($\sim{20}$ pc) is reasonable since temperature mixing is restricted. \citet{ler11} ran both simple modified black body fits and more complex dust models from \citet{dra07} to find $\tau_{160}$ using the $Spitzer$ data for the LMC and SMC and found both produced similar results. A future follow-up study of \citet{gor14} will run more complex dust modeling of the HERITAGE $Herschel$ data.

This study focuses on using dust emission as a means to estimate the amount of molecular gas, which does not require a measurement of the dust mass. By only using $\tau_{160}$ we avoid making any assumptions about the conversion to dust mass, which would introduce a further layer of uncertainty. We define our effective gas-to-dust (\dGDR) ratio in terms of $\tau_{160}$, 
\[\dGDR=\sighi/\tau_{160},\]
such that any proportionality constant between the infrared intensity and $\tau_{160}$ will be incorporated into \dGDR\ and not affect our final results. 

We expect, in principle, that the relationship between \Nhi\ and $\tau_{160}$ should go through the origin, but our measurements show indications of an offset (see Figure \ref{fig:lmc_Nhi_tau}). We regionally fit and then remove the offset and find that the relationship has a positive and roughly constant offset in \Nhi\ in both the LMC ($\Nhi\sim{4}\times10^{20}$ cm$^{-2}$) and SMC ($\Nhi\sim{1.5}\times10^{21}$ cm$^{-2}$). A similar offset is observed by \citet{ler11}, \citet{bol11}, and \citet{rom14}. As opposed to \citet{bol11}, we remove the offset to avoid overestimates when creating maps of the gas-to-dust ratios, which would result in higher estimates of the total amount of gas. This offset could be due to a layer of \hi\ gas with little to no dust, it could be due to the issues with background subtraction with the $Herschel$ images (particularly in the LMC where the HERITAGE maps to not extend much past the main part of the galaxy), or some combination of the two effects. Another possibility is that the relationship between \Nhi\ and $\tau_{160}$ is non-linear and the slope (gas-to-dust ratio) decreases at low \Nhi, which we explore as part of the systematic uncertainty estimation (see Section \ref{subsubsection:uncertainty} and \ref{subsubsection:systematic_unc}). Determining the true nature of the offset is beyond the scope of this work, but warrants further investigation. We subtract the offset in \Nhi\ from the \hi\ map and use the offset-subtracted map for the rest of the analysis. For further discussion on the offset subtraction see Appendix \ref{appendix:hi_offset}.

\vspace{\baselineskip}

\noindent \underline{Steps to Produce Molecular Gas Map}
\begin{enumerate}
\item Model the dust emission in the Herschel images to get $\tau_{160}$ (see Appendix \ref{appendix:modeling_dust} for more details).
\item Fit the \hi\ offset in the \Nhi\ vs. $\tau_{160}$ distribution regionally (see Appendix \ref{appendix:hi_offset} for more details). 
\item Produce first iteration map of the spatially varying effective gas-to-dust ratio ($\dGDR$) at 500 pc scales determined from the diffuse regions ($\siggas=\sighi$)
	\begin{enumerate}
	\item Compute \dGDR\ for each pixel.
	\item Mask all pixels that likely have molecular gas: all regions within 2\arcmin\ of bright CO emission ( $I_{\text{CO}}>3\sigma$)
	\item Use averaging of nearest neighbors to iteratively fill in the masked (molecular) regions in the map.
	\item Convolve map with symmetric Gaussian with FWHM = 500 pc.
	\end{enumerate}
\item Estimate \sigmol\ using the first iteration of the smoothed effective \dGDR:
\[ \sigmol = (\dGDR\sigdust) - \sighi. \]
\item Produce second iteration of map of spatially varying \dGDR\ smoothed to 500 pc. Same as Step 4 with the modification that both regions within 2\arcmin\ of bright CO emission $(I_{\text{CO}}>3\sigma)$ and points that have estimated $\sigmol > 0.5\sighi$ are masked.
\item Produce final map of \sigmol\ map using the second iteration of the smoothed \dGDR\ map. 
\end{enumerate}

The final steps in producing the molecular gas maps remove unphysical artifacts. First, we remove small regions of estimated \htwo\ that are likely spurious by masking pixels that have positive molecular gas in less than 50\% of the pixels surrounding them within a $4\arcmin\times4\arcmin$ box ($12\times12$ pixels in the modified black body map from this work and $4\times4$ pixels in the maps from \citet{gor14}; $\sim60\times60$ pc in the LMC and $\sim70\times70$ pc in the SMC). Generally, this removes emission smaller than $\sim2\arcmin$ ($r\sim{30}$ pc in the LMC and $r\sim{35}$ pc in the SMC)--two times the beam size of the lower resolution \hi\ data--and regions of negative values (from under-estimated total gas). Second, we median-filter the map over 3 pixels ($\sim1\arcmin$ in the LMC map from this work) to smooth out the \sigmol\ map and remove spikes that are unphysical and below the resolution of the \hi\ map, largely due to the residual striping from the HERITAGE PACS images \citep{mei13}. 

There are a few caveats to this methodology that can potentially bias our molecular gas estimate. In addition to tracing the molecular gas (including any ``CO-faint'' component), our methodology may also trace optically thick and/or cold \hi\ gas that emits disproportionately to the optically thin \hi. \citet{sta99} takes a statistical approach and estimates the optical depth correction in the SMC based on column density using the absorption line measurements from \citet{dic00} and finds the correction only changes the total \hi\ mass by $\sim{10\%}$. \citet{lee15b} takes a similar approach to estimate an optical depth correction in the Milky Way and finds that the correction only increases the mass of \hi\ in the Perseus molecular cloud by $\sim{10\%}$. \citet{bra12} attempted to measure the \hi\ optical depth from the flattening of the line profile in M31, M33, and the LMC, and found non-negligible optical depth corrections for high column densities ($22<\text{log \Nhi}<23$) in compact ($\sim{100}$ pc) regions, which increases the total \hi\ mass by $\sim{30\%}$. The \citet{bra12} estimate relies on the assumption of gaussian line profiles to look for flattening of the \hi\ line due to optical depth, which is a difficult measurement in the low signal-to-noise data. \citet{mck15} find $\sim{30\%}$ to be the appropriate \hi\ optical depth correction for the Solar neighborhood based on the average correction factors found using absorption line measurements in the plane of the Milky Way. \citet{fuk15}, on the other hand, find more extreme opacity correction factors, as high as a factor of $\sim{2}$ in the plane of the Milky Way using a relationship between \Nhi\ and the optical depth at 353 GHz from $Planck$. The possible \hi\ opacity corrections coming from a variety of methods and data show that the factors are uncertain.

The manner in which the optical depth correction will affect our molecular gas estimates is complex. It can increase the \hi\ column density in the regions used to estimate the gas-to-dust ratio, leading to an increase in the total gas estimated in the molecular regions, and/or in the molecular regions, resulting in a decrease in the amount of molecular gas. We choose to use the \hi\ statistical opacity corrections from \citet{sta99} and \citet{lee15b} to explore how correcting for optical depth effects our methodology in Section \ref{subsubsection:hi_contribution}.

We note that, in Perseus where the structure of the molecular cloud is resolved, \citet{lee15b} compares their map of \hi\ with the statistical optical depth correction to their inferred ``CO-faint'' gas, observing that the structures are not spatially coincident  (see Figure \ref{fig:lmc_h2_regions} and \citealt{lee15b}). This suggests that the ``CO-faint'' gas cannot be explained by optically thick \hi\ alone. Additionally, \citet{lee15b} comment that their opacity corrected \hi\ map does not show the the sharp peaks seen in maps from \citet{bra12}.   

Our methodology also relies on the assumption that the gas-to-dust ratio in the diffuse, atomic gas is the same in the molecular regions; we only measure the relationship between gas and dust in the atomic phase. There is observational evidence that the gas-to-dust ratio may vary from the diffuse to the dense gas in the Magellanic Clouds \citep{bot04, rom14}. In the Milky Way, $Planck$ results show an factor of 2 increase in the FIR dust optical depth per unit column density ($\tau_{250}/N_{H}$) from the diffuse to the dense gas \citep{pla11b}, which would could indicate a lower gas-to-dust ratio in the dense gas. Both optically thick \hi\ and a decrease in the gas-to-dust ratio from the diffuse to the dense gas would mimic the effect of molecular gas and would result in our methodology overestimating the amount of molecular  gas. We explore how these factors could affect our measurement of \htwo\ in our systematic uncertainty estimate.

\subsubsection{Map Sensitivity and Uncertainty}
\label{subsubsection:uncertainty}

We use a Monte Carlo method to estimate the uncertainty in our molecular gas maps and determine the sensitivity levels. For the maps produced with the dust fitting from this work, we select three sub-regions (shown in Figure \ref{fig:lmc_h2_map}) with different levels of molecular gas (high, moderate, and low). We add normally distributed noise with an amplitude equal to the uncertainty to each of the $Herschel$ bands and fit $T_{d}$ for a fixed $\beta$ for each sub-region and then calculate $\tau_{160}$. For the dust modeling results from \citet{gor14}, we add normally distributed noise to the $\tau_{160}$ maps with amplitude equal to the uncertainty estimates from \citet{gor14}. Finally, we add noise to the \hi\ map and create new \sigmol\ maps. The process is repeated 100 times for each of the different maps. We use the distribution of \sigmol\ for each pixel from the Monte Carlo realizations to estimate a realistic uncertainty. The sensitivity of the maps is estimated by finding the lowest \sigmol\ that is consistently recovered at $\ge2\sigma$. 

We know that the systematic uncertainty from the methodology will dominate the uncertainty in our molecular gas maps \citep{ler09,bol11}. To estimate the level of systematic uncertainty, we see how changes to various aspect of the mapping methodology affect the estimated total molecular mass \Mmol\ (which includes the factor of 1.4 to account for \he). We explore the effects of different assumptions in the dust modeling and determination of the gas-to-dust ratio and produces maps that:
\begin{itemize} 
\item change the value of $\beta$ in our dust modeling and re-run the fitting with $\beta=1.5$ and $\beta=2.0$; 
\item do not remove an \hi\ offset, which explores the idea that the relationship between \Nhi\ and $\tau_{160}$ may not be linear a low column densities;
\item apply a single gas-to-dust ratio using the high and low values from \citet{rom14} (as opposed to using the map of \dGDR);
\item scale the \dGDR\ map down by a factor of 2 in the molecular regions to account for a possible change in the gas-to-dust ration from the diffuse to the dense gas, where we define the dense gas as regions in the map that are likely to have molecular gas (Step 5 in Section \ref{subsection:h2_method});
\item apply a single gas-to-dust ratio for the diffuse gas and a lower value for the dense gas using the values from \citet{rom14}, where the dense gas value is applied to regions with bright CO emission (as in \citealt{rom14}). 
\end{itemize}
For the versions of the maps where we use gas-to-dust ratios found in \cite{rom14}, we use the maps of $\sigdust$ in place of $\tau_{160}$. We use the range in \Mmol\ values to estimate the amount of systematic uncertainty in our molecular gas estimate.

\subsubsection{Estimating \htwo\ from CO}
\label{subsection:CO_H2}

For the purposes of this work, we want to compare the amount of \htwo\ traced by detected, bright \co\ emission to the molecular gas traced by the dust emission. To convert the CO intensity ($I_{\mbox{\scriptsize{CO}}}$) into column density of mass, we use the following equations:
\begin{equation}
N(\mbox{H}_{2}) = X_{\mbox{\scriptsize{CO}}}~I_{\mbox{\scriptsize{CO}}}
\end{equation}
\begin{equation}
M_{\mbox{\scriptsize{mol}}}=\alpha_{\mbox{\scriptsize{CO}}}~L_{\mbox{\scriptsize{CO}}},
\end{equation}
where proportionality constants appropriate for Galactic gas are $X_{\mbox{\scriptsize{CO}}}=2\times10^{20}$ cm$^{-2}$ (K km s$^{-1})^{-1}$ and $\alpha_{\mbox{\scriptsize{CO}}}=4.3~\msol$ (K km s$^{-1}$ pc$^{2})^{-1}$, $I_{\mbox{\scriptsize{CO}}}$ is the integrated intensity of the $^{12}$CO $J=1\rightarrow0$ transition (in K km s$^{-1}$), and $L_{\mbox{\scriptsize{CO}}}$ is the luminosity of the same transition (in K km s$^{-1}$ pc$^{2}$). On small spatial scales and in CO-bright regions, using the Galactic values is a good approximation \citep{bol08}.

\subsection{Tracing Recent Star Formation}

We use \ha, locally corrected for extinction using 24 \micron\ emission, to trace the star formation rate surface density ($\Sigma_{\mbox{\scriptsize{SFR}}}$). Following \citet{bol11}, we use the star formation rate (SFR) calibration by \citet{cal07} to convert \ha\ and 24 \micron\ luminosities: 
\begin{align}
\mbox{SFR}(\smpy) & =  5.3 \times 10^{-42}[L(\mbox{H}\alpha) \nonumber \\
 & +~(0.031 \pm 0.006)L(24\micron)],  
\end{align}
where luminosities are in erg s$^{-1}$ and $L(24\micron)$ is expressed as $\nu L(\nu)$. The average contribution from 24 \micron\ to the total star formation rate is $\sim{20\%}$ in the LMC and $\sim{10\%}$ in the SMC. A significant fraction ($\sim{40\%}$) of the \ha\ emission in both the LMC and SMC is diffuse. We include all of the \ha\ emission in this analysis since \citet{pel12} showed that all of the ionizing photons could have originated from \hii\ regions from massive stars (see Appendix \ref{appendix:diffuse_ha} for further discussion). The RMS background value of the SFR map is $1\times10^{-4}$ \smkpc\ in the LMC and $4\times10^{-4}$ \smkpc\ in the SMC. 

This conversion to star formation rate assumes an underlying broken power-law Kroupa initial mass function (IMF) and was calibrated against Paschen-$\alpha$ emission for individual star-forming regions. Ideally, \ha\ and 24 \micron\ emission would only be used for size scales that fully sample the IMF and sustain star formation for $>{10}$ Myr; for smaller scales, pre-main sequence stars are more appropriate and a better indicator of the current star formation rate. \citet{hon15} find that the star formation rate from pre-main sequence stars matches that from \ha\ at scales of $\sim{150}$ pc in the N66 region in the SMC. Our highest resolution of $\sim{20}$ pc resolves \hii\ regions, and the mapping of the star formation rate on these scales is questionable. Nonetheless, we apply the star formation rate conversion even to our highest resolution data to allow us to compare to other studies and investigate the relationships in terms of a physical quantity, although it is important to keep these limitations in mind when interpreting the results.

\subsection{Convolving to Lower Resolutions}

To produce the lower resolution molecular gas maps, we first convolve the maps from the $Herschel$ beam to a gaussian with FWHM of $30\arcsec$ for the $\beta=1.8$ map (appropriate for the 350 \micron\ image resolution) and $40\arcsec$ for the BEMBB map (appropriate for the 500 \micron\ image resolution) using the kernels from \citet{ani11}. We then produce the range of lower resolution maps (from 20 pc to $\sim{1}$ kpc) by convolving the images of \sigsfr, \sigmol, and \sighi\ with a gaussian kernel with FWHM $ = \sqrt{(r^{2} - r_{0}^{2})}$, where $r$ is the desired resolution and $r_{0}$ is the starting resolution of the image. The images are then resampled to have approximately independent pixels (one pixel per resolution element). To mitigate edge effects from the convolution, we remove the outer two pixels (two beams) for all resolution images of the LMC. In the SMC, we remove two outer pixels for $r\leq600$ pc and remove one pixel from the edges for $r\geq700$ pc due to the small size of the images. 

\section{Results}
\label{section:results}

\input{table1.tex}

\input{table2.tex}
\subsection{Molecular Gas in the Magellanic Clouds}
\begin{figure*}[t]
\epsscale{1.0}
\plotone{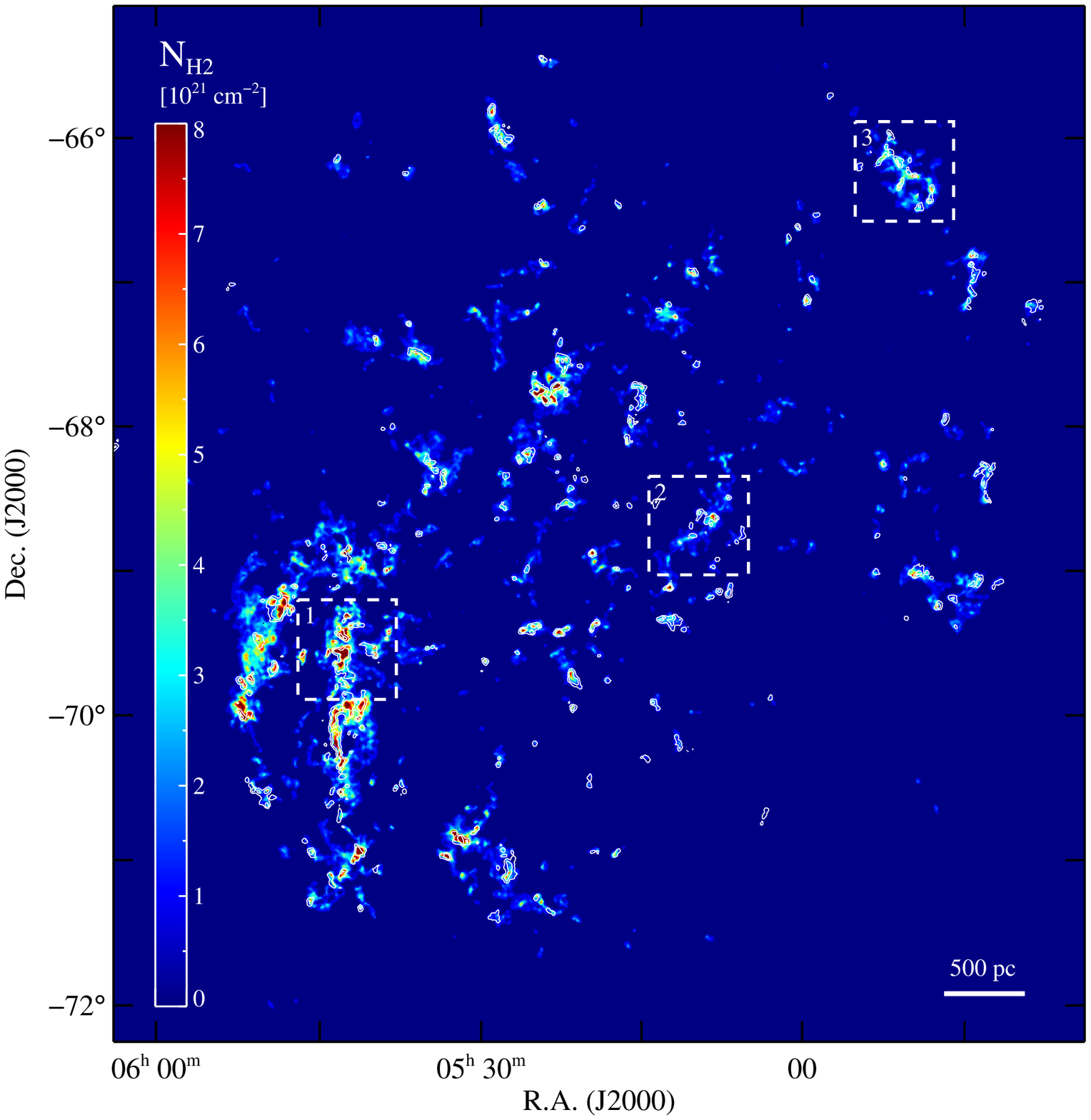}
\caption{\htwo\ column density (\Nhtwo) map of the LMC at $\sim{5}$ pc resolution ($\theta=20\arcsec$, 1 beam per pixel sampling) produced by modeling the dust continuum emission from $Herschel$ 100 \micron, 160 \micron, 250 \micron, and 350 \micron\ observations from HERITAGE \citep{mei13} using a modified black body. The white contours show the 1.2 K km s$^{-1}$ ($3\sigma$) and 5 K km s$^{-1}$ levels of the MAGMA DR3 CO map ($\theta=40\arcsec$), which covered regions with prior CO detection. Assuming a Galactic conversion factor of $X_\textnormal{CO} = 2\times10^{20}$ cm$^{-2}$ (K km s$^{-1})^{-1}$, the contour levels correspond to column densities of $2.4\times10^{20}$ cm$^{-2}$ and $1\times10^{21}$ cm$^{-2}$. The dashed white boxes indicate the three regions in Figure \ref{fig:lmc_h2_regions}. There is excellent agreement between the dust-based molecular gas map and the CO map even though the CO is not directly used to produce the map.
\label{fig:lmc_h2_map}}
\end{figure*}

\begin{figure*}[t]
\epsscale{0.8}
\plotone{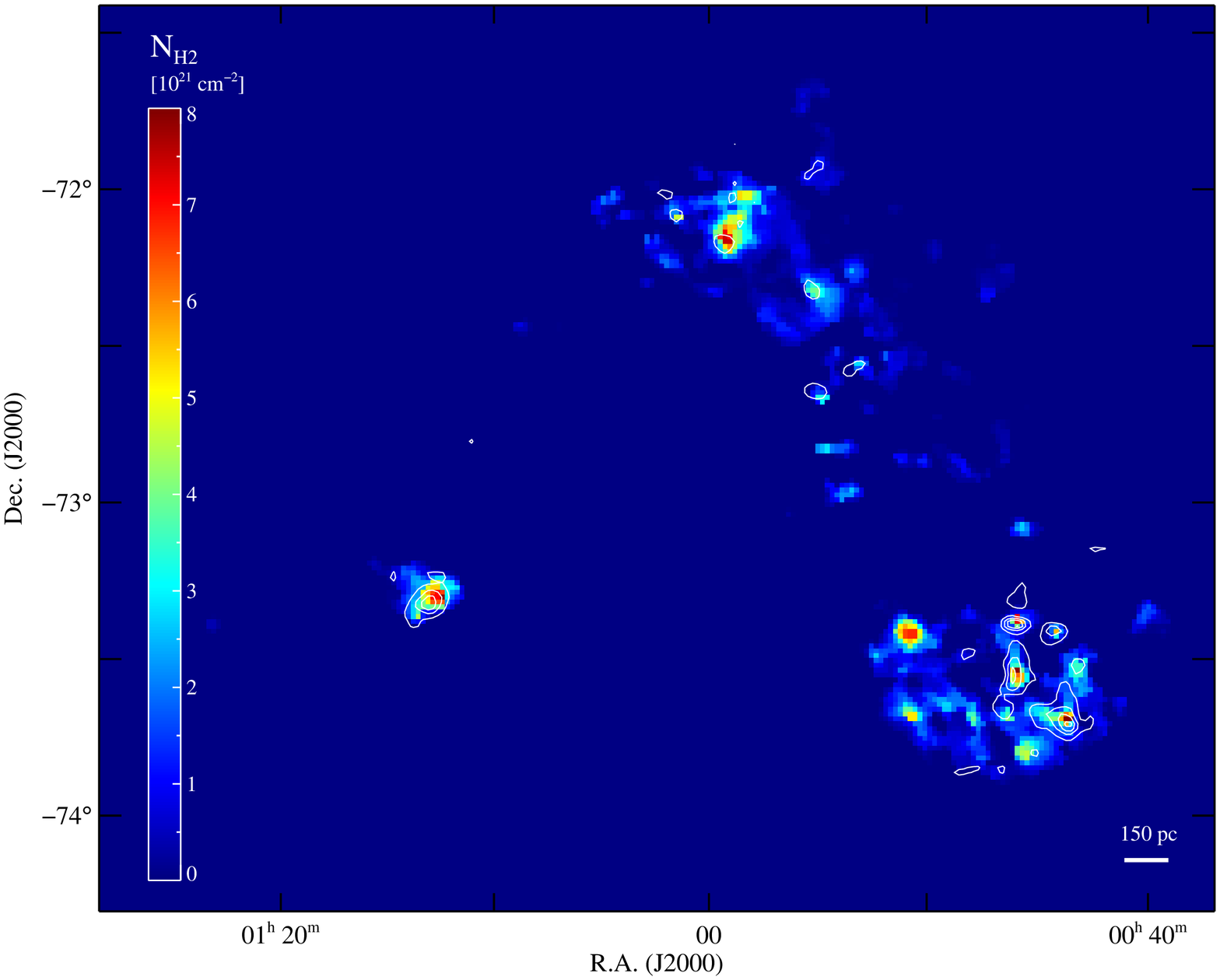}
\caption{\htwo\ column density (\Nhtwo) map of the SMC at $\sim{10}$ pc resolution ($\theta=40\arcsec$, 1 beam per pixel sampling) produced by modeling the dust continuum emission from $Herschel$ 100 \micron, 160 \micron, 250 \micron, 350 \micron, and 500 \micron\ observations from HERITAGE \citep{mei13} using the BEMBB dust modeling results from \citet{gor14}. The white contours show the 0.45 ($3\sigma$), 1, 1.5, and 2 K km s$^{-1}$ levels of the NANTEN CO map ($\theta=1\arcmin$).
\label{fig:smc_h2_map}}
\end{figure*}

\begin{figure*}[t]
\epsscale{1.2}
\plotone{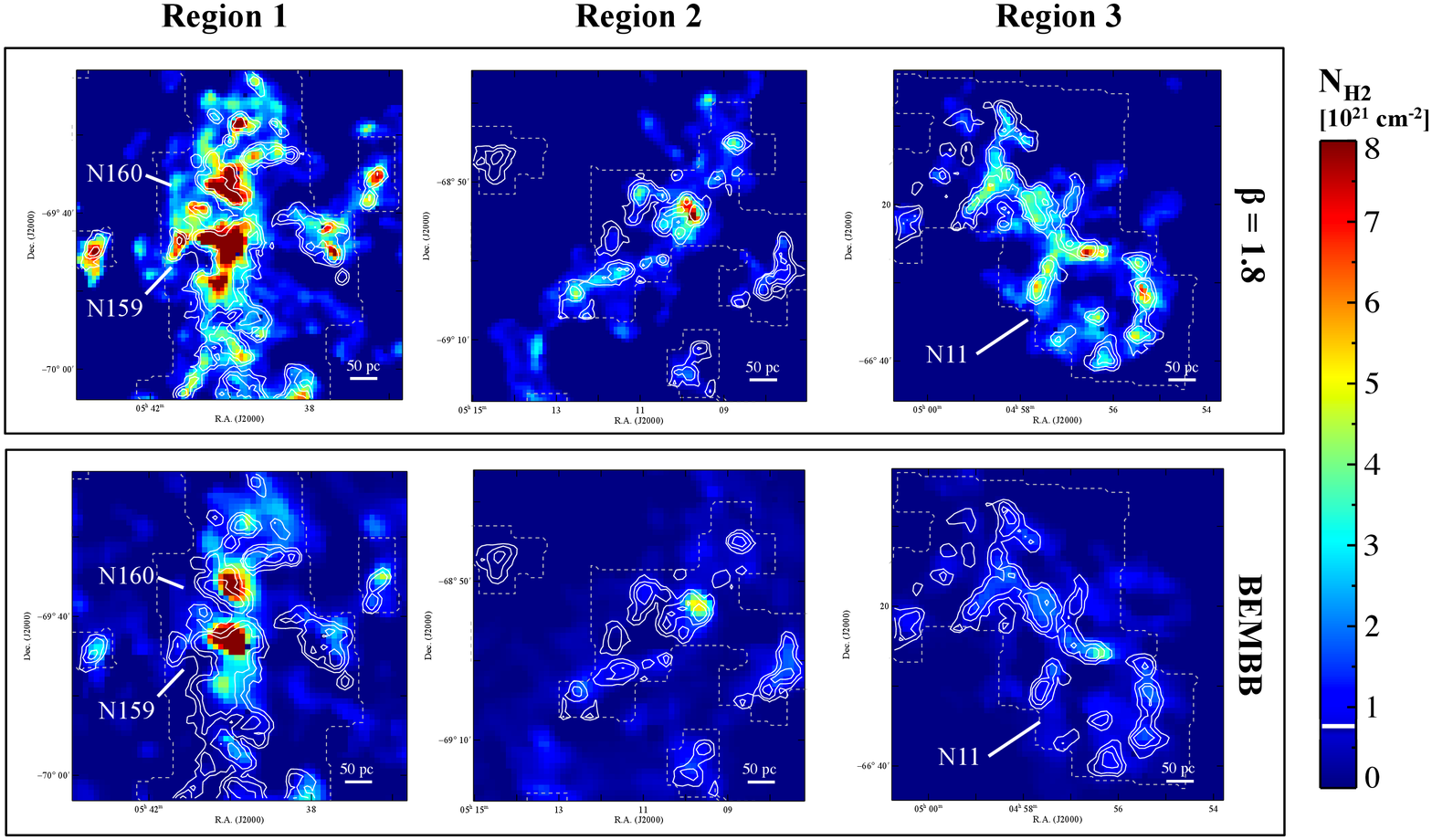}
\caption{The top and bottom rows of images respectively show the enlarged regions of the \Nmol\ maps (identified in Figure \ref{fig:lmc_h2_map}) for the dust modeling with $\beta=1.8$ and the BEMBB model from \citet{gor14} at the same color scale as show in Figure \ref{fig:lmc_h2_map}. The contours show the MAGMA $^{12}$CO intensity at levels of 0.6 ($3\sigma$), 2, and 5 K km s$^{-1}$ with the dashed grey line showing the survey coverage in the regions. The white line on the color bar indicates the estimated sensitivity level of $\Nmol\sim{7\times10^{20}}$ cm$^{-2}$ ($\sigmol\sim{15}$ \smpc). Both dust-based molecular gas maps show similar structure. The dust-based estimate tends to show more extended molecular gas than that traced by $^{12}$CO. The only clear example of a CO cloud (with strong CO emission) with no dust-based molecular gas counterpart (in both the LMC and SMC) is found in the NE of Region 2. The difference in intensity demonstrates the systematic uncertainty in the methodology. 
\label{fig:lmc_h2_regions}}
\end{figure*}

We find molecular gas fractions that are comparable to the Milky Way in the LMC ($17\%$), but much lower in the SMC ($3\%$). These molecular gas fractions come from our new estimates of the total molecular gas mass: we find a total molecular gas mass (including \he) in the LMC of $M_{\mathrm{LMC}}^{\mathrm{mol}} = 6.3 ^{+6.3}_{-3.2} \times 10^{7}$ \msol\ and $M_{\mathrm{SMC}}^{\mathrm{mol}} = 1.3 ^{+1.3}_{-0.65}\times 10^{7}$ \msol\ in the SMC. These values are the sums (with no cuts) of our fiducial molecular gas maps that use the BEMBB dust modeling results from \citet{gor14} with a spatially varying \dGDR\ and include a factor of 2 systematic uncertainty. Table \ref{table:h2_maps} shows the results from our exploration of varying the map making methodology to estimate the systematic uncertainty combined with estimates of the molecular gas mass from the literature. In Table \ref{table:integrated_properties} we list the integrated properties of both galaxies. The molecular gas maps are sensitive to $\sigmol\sim{15}$ \smpc\ ($\sim{7\times10^{20}}$ cm$^{-2}$) based on the Monte Carlo estimates, which is comparable to the sensitivity of the SMC map from \citet{bol11}. Our molecular gas fraction in the SMC is lower than previous estimates \citep{ler07,bol11}, but is consistent with the factor of $\sim{2}$ estimate of systematic uncertainty for all of the estimates (see Appendix \ref{appendix:prev_work} for further discussion). 

The fiducial molecular gas maps ($\beta=1.8$ and BEMBB dust modeling) were produced using maps of the effective dust-to-gas ratio (\dGDR) that had average values of $\Nhi/\tau_{160}$ of $1.8\pm0.6\times10^{25}$ cm$^{-2}$ (LMC $\beta=1.8$), $1.3\pm0.3\times10^{25}$ cm$^{-2}$ (LMC BEMBB), and $4.8\pm0.9\times10^{25}$ cm$^{-2}$ (SMC BEMBB). In the Milky Way, \citet{pla14} found $\Nhi/\tau_{160} = 1.1\times10^{25}$ cm$^{-2}$ in the diffuse ISM. Our $\Nhi/\tau_{160}$ values are on average a factor of $\sim{1.5}$ (LMC) and $\sim{4.4}$ (SMC) times higher than the diffuse Milky Way ISM, which is consistent with the expectation that the gas-to-dust ratio should increase with decreasing metallicity. 

Given the total NANTEN CO luminosities of $L(\text{CO})_{\text{LMC}}=7\times10^{6}$ K km s$^{-1}$ pc$^{2}$ and $L(\text{CO})_{\text{SMC}}=1.7\times10^{5}$ K km s$^{-1}$ pc$^{2}$ (using no sensitivity cuts), we find $\alpha^{\text{LMC}}_\text{CO}=10^{+9}_{-6}$ and $\alpha^{\text{SMC}}_\text{CO}=76^{+77}_{-38}$, where all units for $\alpha_{\text{CO}}$ are given in \msol\ (K km s$^{-1}$ pc$^{2}$)$^{-1}$. Compared to the Milky Way value of $\alpha_{\text{CO}}=4.3$ \citep{bol13}, the conversion factor for the LMC is $\sim{2}$ times higher and the SMC is $\sim{17}$ times higher. Our $\alpha_{\text{CO}}$ values are comparable to the dust-based $\alpha_{\text{CO}}$ found by \citet{ler11} of $6.6$ (K km s$^{-1}$ pc$^{2}$)$^{-1}$ and $53-85$ (K km s$^{-1}$ pc$^{2}$)$^{-1}$ for the LMC and SMC, respectively.

\subsubsection{Structure of the Molecular Gas}

One of the most striking results is the similarity of the structure of the molecular gas traced by dust to that traced by CO  throughout the entire LMC (see Figure \ref{fig:lmc_h2_map} and Figure \ref{fig:lmc_h2_regions}). Since our methodology only indirectly uses the CO map as a mask (see Section \ref{subsection:h2_method}) the similarity is confirmation that our methodology traces the structure of the gas. Figure \ref{fig:lmc_h2_regions} shows that both dust modeling techniques produce maps with similar structure, although the BEMBB map tends to predict systematically lower amounts of \htwo. 

The details of the structures traced by CO are different from the dust-based molecular gas map. All of the regions shown in Figure \ref{fig:lmc_h2_regions} show molecular gas traced by dust, but not by CO at the $3\sigma$ level. This is likely a layer of self-shielded \htwo\ where CO has mostly dissociated, as expected from models \citep{wol10,glo11}. The same is generally true for the SMC, but having only the lower resolution full coverage NANTEN \co\ map ($r=2.6\arcmin$) makes detailed comparison of the structure difficult. Conversely, Region 2 in Figure \ref{fig:lmc_h2_regions} shows a molecular gas cloud traced by CO and not by the dust-based method. As discussed in \citet{ler09}, one possible explanation is that the dust is cold and faintly emitting in the far infrared, below the sensitivity of the HERITAGE $Herschel$ images. The peak in the CO emission of this cloud is detected from 250-500 \micron, but only weakly detected at 160 \micron\ and marginally detected ($\sim{3\sigma}$) at 100 \micron, consistent with the interpretation of cold dust. There are a few other detections of CO without a dust-based molecular gas counterpart, although the cloud in Region 2 is the clearest example with the strongest CO emission.

\subsubsection{Systematic Uncertainty}
\label{subsubsection:systematic_unc}

The systematic uncertainty comes from the different possible assumptions that can be made in the dust modeling and the method of measuring the gas-to-dust ratio. Because the statistical errors are typically small, the systematic uncertainty dominates the total uncertainty in the molecular gas mapping methodology \citep{ler09,bol11}. We present the range of our total molecular gas mass estimates (\Mmol) in Table \ref{table:h2_maps} (we list all \Mmol\ estimates alongside estimates from the literature in Table \ref{table:h2_estimates} in Appendix \ref{appendix:prev_work}). We use the range of  as a means to gauge the amount of total systematic uncertainty and we look at the variation between the two fiducial molecular gas maps (Table \ref{table:h2_maps} rows 1 and 2) with different dust modeling assumptions to determine the amount of systematic uncertainty due to assumptions in the dust modeling.  

The two lowest \Mmol\ estimates that we found use a single gas-to-dust ratio of 380 from \citet{rom14} and use the upper estimates for the gas-to-dust ratios for the diffuse and the dense gas from \citet{rom14} of GDR$_{\text{diffuse}}=540$ and GDR$_{\text{dense}}=330$ (see rows 6 and 10 in Table \ref{table:h2_estimates} in Appendix \ref{appendix:prev_work}). These maps have large regions of negative values from where the estimated total gas is less than the \hi, which causes only small areas of estimated \htwo\ and the low \Mmol\ values, which would be due to using too low a value of the gas-to-dust ratio. The value for \Mmol\ with GDR = 380 is less than the total molecular gas you would get by applying a Galactic CO-to-\htwo\ conversion factor to the low resolution NANTEN CO map, which is a lower limit on the total molecular gas since a higher conversion factor should be appropriate when the CO structure is unresolved. We do not consider these values of \Mmol\ when estimating the systematic uncertainty in the total molecular gas mass. 

The difference between the highest (row 5) and lowest (row 6) molecular gas mass is $\sim{3.5}$. The minimum \Mmol\ estimate (row 6) comes from assuming a single gas-to-dust ratio of 540, which is the highest value found by \citet{rom14}. This \Mmol\ estimate is only a factor of $\sim{1.5}$ lower than using a spatially varying \dGDR\ applied to the same BEMBB dust modeling results. The maximum value comes from using the BEMBB modeling that does not remove an \hi\ offset (row 5), which allows for a possible non-linear relationship in \Nhi\ vs. $\tau_{160}$ (see Section \ref{subsubsection:uncertainty} and Appendix \ref{appendix:hi_offset}). This would be an overestimate if the relationship between \Nhi\ vs. $\tau_{160}$ is linear since it will artificially increase the \dGDR\ values in the maps. Allowing for a difference in the gas-to-dust ratio in the diffuse and dense gas by scaling down \dGDR\ in the dense gas reduces \Mmol\ by a factor of $\sim{2}$ for the $\beta=1.8$ map and $\sim{1.5}$ for the BEMBB map. We conclude that our molecular gas estimate is good to within a factor of $\sim{2}$, which agrees with the estimates from similar methodologies by \citet{ler09,ler11} and \citet{bol11}. 

We compare the effects of different dust modeling techniques by using $\beta=1.8$ and BEMBB maps while keeping all other aspects of the methodology the same (using a spatially varying \dGDR). Figure \ref{fig:lmc_h2_regions} shows the difference between the molecular gas maps using the $\beta=1.8$ and BEMBB modeling (top and bottom rows, respectively). The BEMBB map is a factor of $\sim{2}$ lower molecular gas column density estimates than using the fits from the $\beta=1.8$ model. The difference in values between the two maps show no variation as a function of $\tau_{160}$, which indicates that the dust models do not produce systematically different results in the dense gas as compared to the diffuse. The $\tau_{160}$ values from the BEMBB modeling \citep{gor14} tend to be higher than from the $\beta=1.8$ modeling largely due to differences in the fitted dust temperatures ($T_{d}$). The BEMBB modeling tends to fit higher $T_{d}$, which is a result of the range of $\beta$ values combined with the degeneracy between $\beta$ and $T_{d}$ \citep{dup03,she09}: fitting a lower $\beta$ value to the same data will result in an increase in $T_{d}$. An increase in $\tau_{160}$ produces lower effective gas-to-dust ratio (\dGDR) and a lower estimate of the amount molecular gas. Our adopted factor of $\sim{2}$ systematic uncertainty is consistent with the variation seen between the two maps.

\subsubsection{Estimating the Effect of the Optical Depth of \hi}
\label{subsubsection:hi_contribution}

We apply the statistical optical depth corrections from \citet{sta99} for the SMC and \citet{lee15b} for the Milky Way to \Nhi\ maps to estimate how accounting for optically thick \hi\ could affect our molecular gas estimate. There is no comparable optical depth correction for the LMC, so we apply the statistical corrections for the lower metallicity SMC and higher metallicity Milky Way to estimate a range of possible effects. Applying the \citet{lee15b} correction to the LMC \hi\ produces a maximum correction factor of 1.43 and shifts the top $5\%$ of \Nhi\ from $>3.1\times10^{21}$ cm$^{-2}$ to $>4.0\times10^{21}$ cm$^{-2}$, whereas the \citet{sta99} correction produces a maximum correction factor of 1.36 and shifts the top $5\%$ to $>3.3\times10^{21}$ cm$^{-2}$. Applying the \citet{sta99} correction to the SMC produces a maximum correction factor of 1.48 and shifts the top $5\%$ of the \Nhi\ from $>4.4\times10^{21}$ cm$^{-2}$ to $>5.2\times10^{21}$ cm$^{-2}$. In the LMC, both of the statistical optical depth corrections decrease the total molecular gas mass estimate by $\sim{5\%}$ while in the SMC it increases the total molecular gas mass by a factor $\sim{2}$, both are within our estimate of the systematic uncertainty. 

The molecular gas estimate changes because the amount of \hi\ in the diffuse regions, where we determine the effective gas-to-dust ratio, increases or decreases with respect to the amount of \hi\ in the molecular regions. Optical depth corrections in the diffuse gas will increase the effective gas-to-dust ratio and increase the estimate of the total amount of gas. If the optical depth corrections in the molecular regions are similar to the corrections in the diffuse regions, as is the case in the SMC, the total amount of gas will increase and the molecular gas estimate will increase. If the optical depth corrections in the molecular regions are larger than in the diffuse, more of the total gas estimate will be due to \hi\ as opposed to \htwo\ and the molecular gas estimate will decrease, as is the case in the LMC. Ultimately, the decrease in the molecular gas mass estimate in the LMC is negligible, which indicates that optical depth effects do not significantly contribute to our molecular gas estimate.

\subsubsection{Comparison to Previous Work}

The molecular gas maps we present are improvements upon previous dust-based \htwo\ estimates given the availability of the $Herschel$ data with increased sensitivity and coverage of the far-infrared combined with improvements in the methodology and more extensive estimation of the systematic uncertainty. Table \ref{table:h2_maps} includes the existing dust-based molecular gas mass estimates for LMC and SMC from the literature. While some of the total molecular mass values are out of the range of the estimate from this work, they can all be reconciled and explained by differences in methodology and limitations in the data. For a more detailed explanation of the differences in the \Mmol\ estimates from previous works see Appendix \ref{appendix:prev_work}.

\subsection{Molecular Gas and Star Formation} 
\begin{figure*}
\epsscale{1.1}
\plottwo{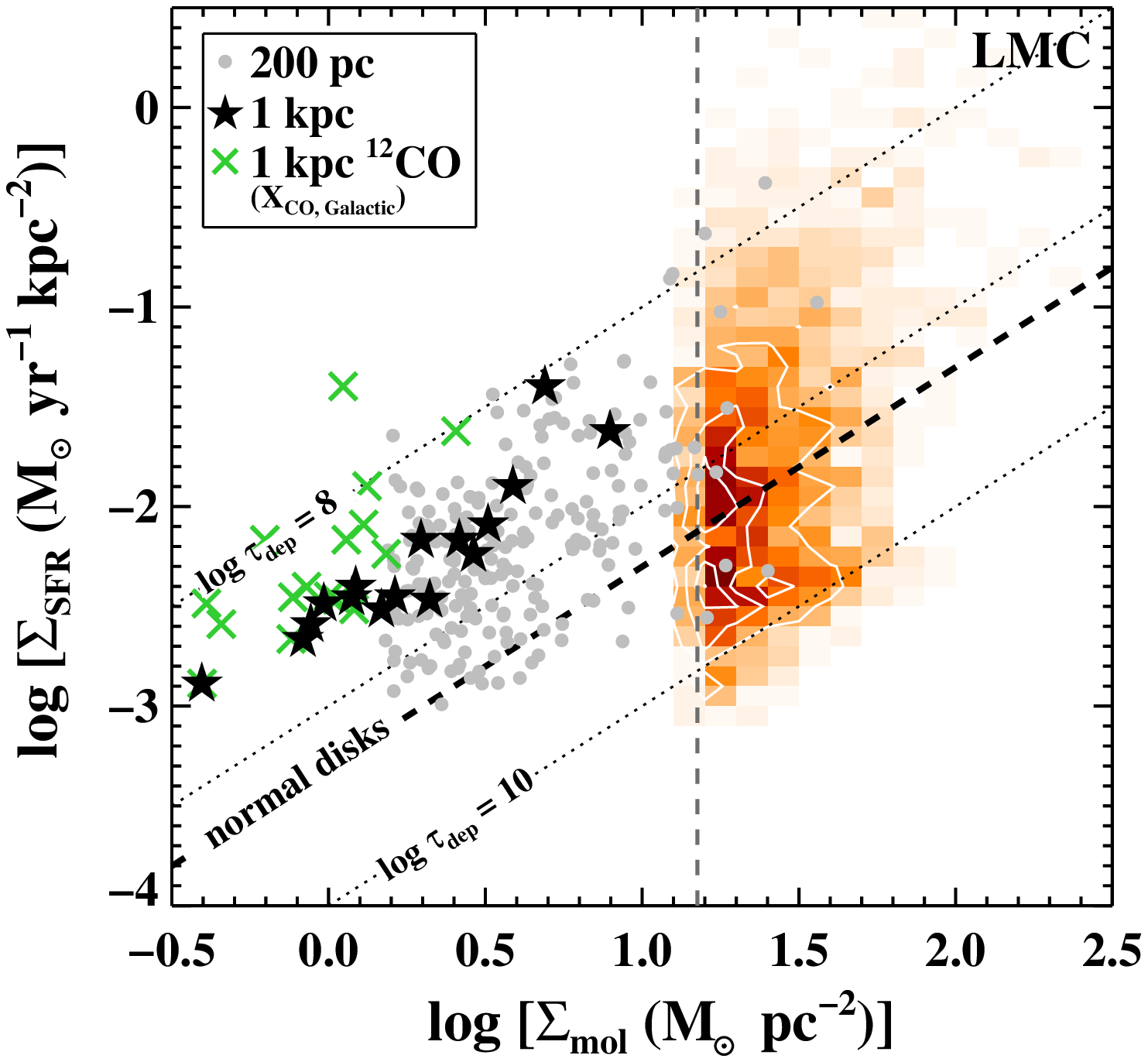}{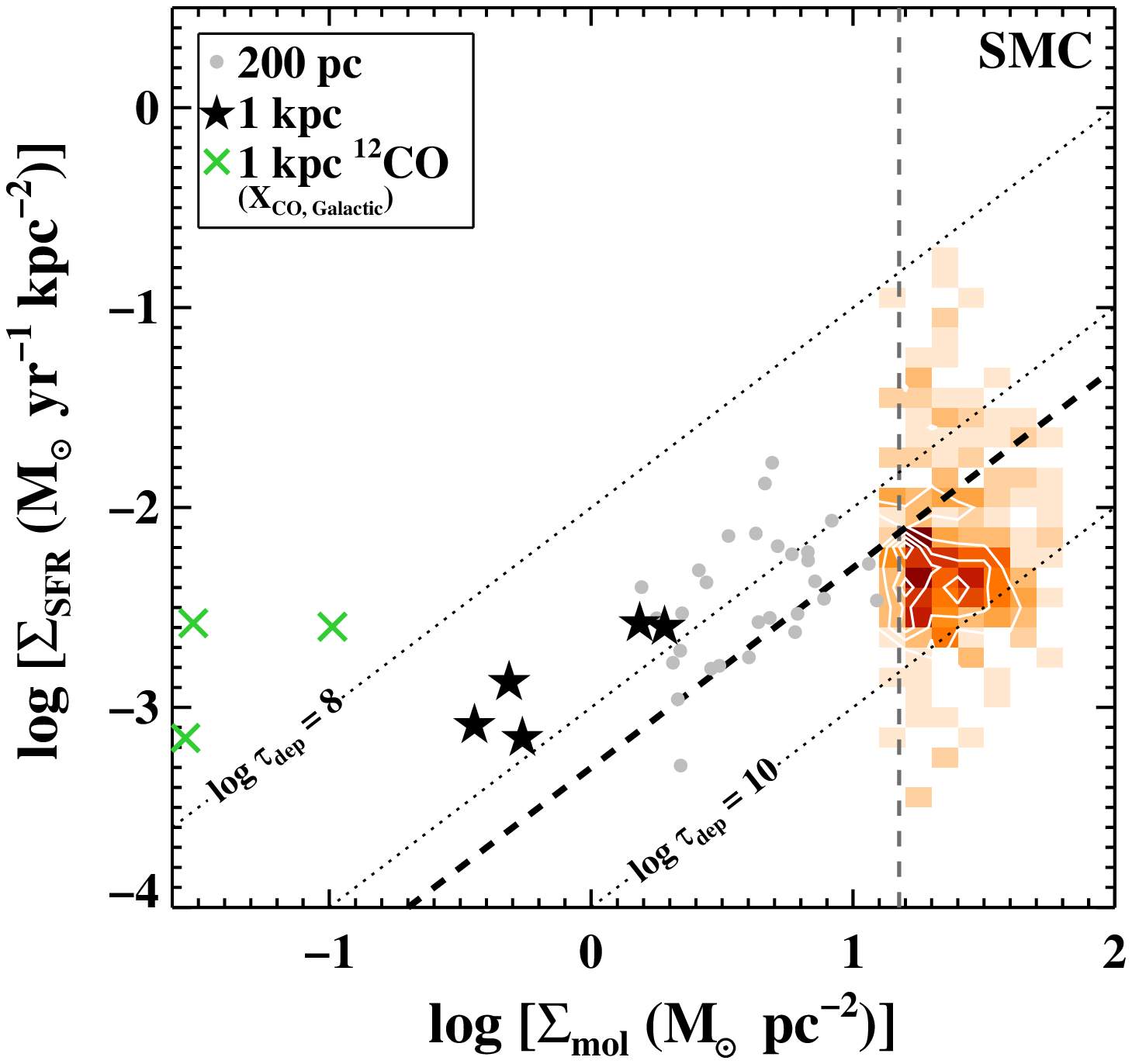}
\caption{\sigsfr\ as a function of \sigmol\ for the LMC (left) and SMC (right) at various resolutions. The red color scale shows the two-dimensional distribution at a resolution of $r=20$ pc with the white contours indicate levels that are 20\%, 40\%, 60\%, and 80\% of the maximum density of points. The vertical gray dashed line indicates the estimated $2\sigma$ sensitivity cut of the $r=20$ pc data ($\sigmol\sim15~\smpc$). The grey circles and black stars show the data at resolutions $r=200$ pc and $r=1$ kpc, respectively. The green stars show \sigmol\ derived from NANTEN CO data at a resolute of $r=1$ kpc using a Galactic CO-to-H$_{2}$ conversion factor. Here we present the SMC data corrected by a higher inclination angle of $i=70\degr$, as opposed to results from \citet{bol11} that used $i=40\degr$, which results in a diagonal shift to lower surface densities.The dotted lines indicate constant molecular depletion times $\taumoldep=0.1$, 1, and 10 Gyr. The dashed line shows the typical depletion time for normal galaxies $\taumoldep\sim{2}$ Gyr \citep{big08,big11,rah12,ler13a}. \label{fig:mol_sfl}}
\end{figure*}
Understanding whether or not metallicity and galaxy mass affect the conversion of gas into stars is important for understanding galaxy evolution throughout cosmic time. The relationship between molecular gas and star formation rate has been studied extensively in nearby, high-metallicity, star-forming galaxies. With the dust-based molecular gas estimates of the nearby Magellanic Clouds, we are in a unique position to probe how the relationship between the molecular gas and star formation rate behave as a function of metallicity and the size scale considered. Figure \ref{fig:mol_sfl} shows the relationships for the LMC and SMC using the new dust-based molecular map at the highest resolution of 20 pc, 200 pc (scale where multiple star forming regions are being averaged), and 1 kpc (comparable to the \co\ surveys of nearby galaxies). 

\begin{figure}
\epsscale{1.2}
\plotone{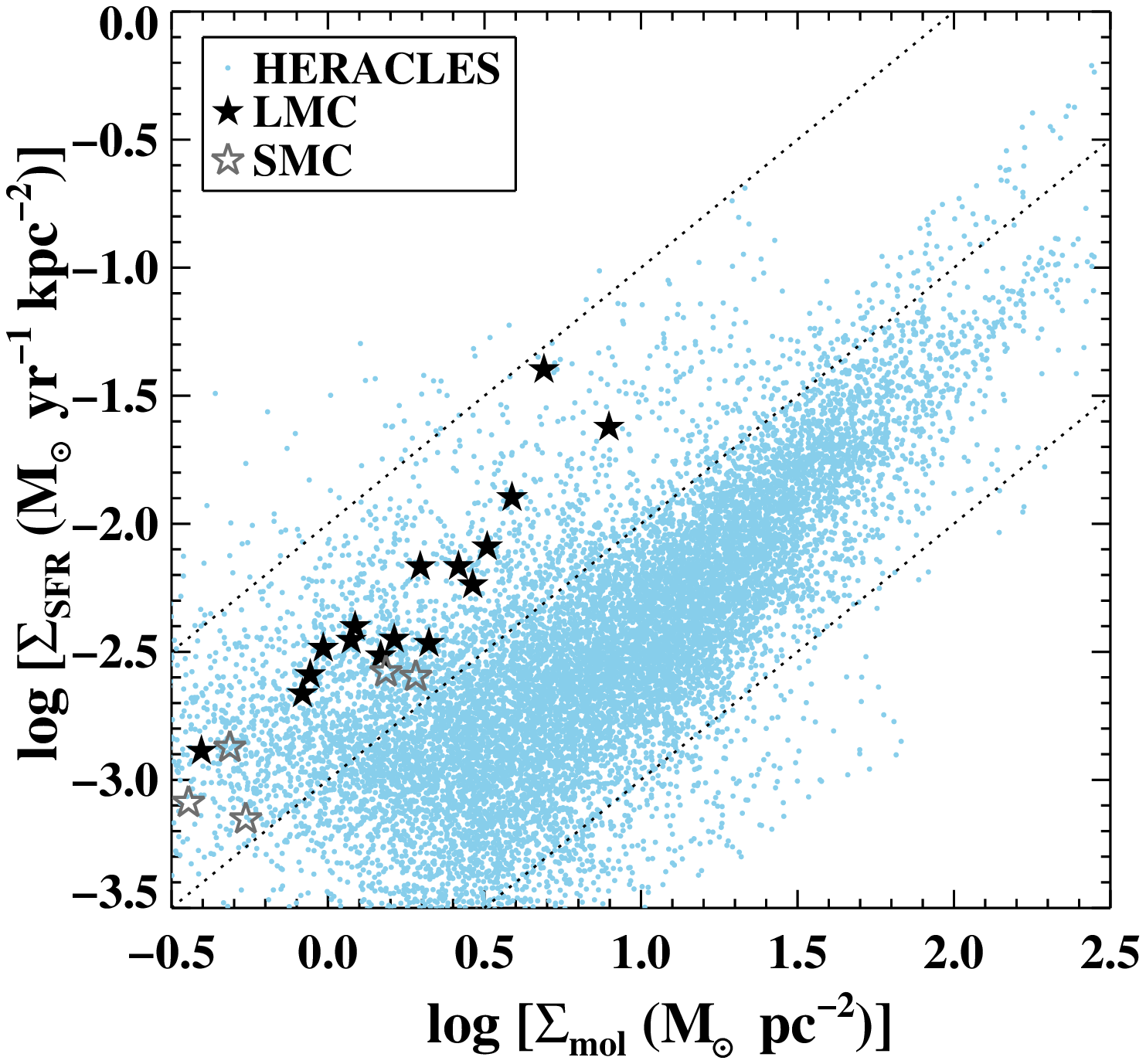}
\caption{\sigsfr\ vs. \sigmol\ for the $r\sim1$ kpc data from the HERACLES sample of nearby star-forming galaxies \citep{ler13a} (blue), where the \sigmol\ is estimated using \co\ with a Galactic CO-to-\htwo\ conversion factor. The $r\sim1$ kpc data for the LMC (filled stars) and SMC (open stars) are over plotted. The LMC and SMC points fall within the full distribution for the HERACLES sample, but offset above the main distribution. 
\label{fig:leroy13_sfl_h2}}
\end{figure}

We compare the relationship between the molecular gas and star formation rate in the SMC and LMC to that for the HERACLES sample of nearby galaxies by \citet{ler13a}. The HERACLES sample resolves the galaxies and compares the gas and star formation at a resolution of $\sim{1}$ kpc. Figure \ref{fig:leroy13_sfl_h2} shows that the LMC and SMC data (convolved to a comparable resolution of 1 kpc) lie within the scatter in the data for high-metallicity, star-forming galaxies, although above the main cluster of data points for a given molecular gas surface density. 

\subsubsection{Molecular Gas Depletion Time}

A convenient way to quantify the relationship between molecular gas and star formation is in terms of the amount of time it would take to deplete the current reservoir of gas given the current rate of star formation, the molecular gas depletion time:
\begin{equation}
\taumoldep=\sigmol/\sigsfr .
\end{equation}
The data for the LMC and SMC appear consistent with a well-defined depletion time. We find average molecular gas depletion times at 1 kpc scales of $\sim{0.4}$ Gyr in the LMC and $\sim{0.6}$ Gyr in the SMC. Weighting the average of \taumoldep\ by the molecular gas mass and star formation rate does not significantly affect the averages at 1 kpc scales; at 200 pc scales, weighting of the average typically changes the value of \taumoldep\ by $\sim{20\%}$. The main exception is for the star formation rate weighted \taumoldep\ average in the LMC, which is shorter by $\sim{50\%}$ and likely due to the significant contribution of 30 Doradus at these scales.  The range of possible molecular gas depletion times given the factor of up to $\sim{2}$ systematic uncertainty in the molecular gas estimate is $\sim{0.2-1.2}$ Gyr. This is shorter than the molecular gas depletion time found for the SMC by \citet{bol11} of $\taumoldep \sim{1.6}$ Gyr at 1 kpc resolution, but within the factor of 2 systematic uncertainty on both estimates. The molecular gas depletion time found in the Magellanic Clouds is lower than the average value of $\sim{2}$ Gyr for nearby normal disk galaxies at comparable $\sim{1}$ kpc size scales, but within the range of observed values for the STING sample \citep{rah12} and the larger HERACLES sample \citep{big08,big11,ler13a}.

\begin{figure}[t]
\epsscale{1.2}
\plotone{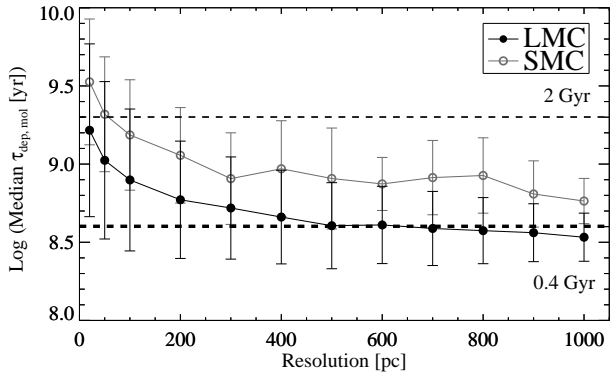}
\caption{Median molecular gas depletion time as a function of resolution. Black filled and open grey circles show the data for the LMC and SMC, respectively. The error bars show $1\sigma$ on the mean. The upper dashed line shows $\taumoldep=2$ Gyr, the average for normal galaxies, and the lower dashed line shows $\taumoldep=0.4$ Gyr, the integrated depletion time for both the LMC and SMC. The LMC and SMC \taumoldep\ reach the integrated value of $\sim0.4$ Gyr and $\sim0.6$ Gyr, respectively, at large ($>500$ pc) scales.
\label{fig:dep_res}}
\end{figure}

Figure \ref{fig:dep_res} shows that the median \taumoldep\ is $\sim{2-3}$ Gyr at the highest resolution of 20 pc. The molecular gas depletion time changes with resolution because the peaks in the molecular gas are physically separated from the peaks in the star formation rate at scales where the star-forming regions are spatially resolved. The tendency of low star formation rates at the peaks in the molecular gas and low to no molecular gas at the peaks in the star formation rate (\taumoldep\ is only defined for regions with \sigmol) biases \taumoldep\ at high resolutions biased towards longer times. A scale of 200 pc is typically large enough to include both the recent star formation and the molecular gas and sample star-forming regions at a range of evolution stages \citep{sch11}. While the molecular gas depletion time gets closer to the integrated \taumoldep\ value at a scale of 200 pc, the median \taumoldep\ reaches the integrated value at $\sim{500}$ pc in the LMC and SMC.

The lower metallicities of the SMC and LMC and the lack of a metallicity bias in our dust-based molecular gas estimate allow us to investigate whether there is any trend in \taumoldep\ with metallicity. Figure \ref{fig:dep_metallicity} shows that there is no clear trend in the average molecular gas depletion times when comparing the LMC and SMC to the HERACLES sample of galaxies. \citet{ler13a} also saw no trend with metallicity as long as they allowed for a variable CO-to-\htwo\ conversion factor. We also compare our measurements to the integrated \taumoldep\ using a metallicity dependent CO-to-\htwo\ conversion factor for the $Herschel$ Dwarf Galaxy Survey (DGS; \citealt{cor14}). Over the range of metallicities studied, the main cause of variations in the molecular gas depletion time does not appear to be metallicity.

\subsection{Correlation Between Gas and Star Formation Rate from 20 pc to 1 kpc Size Scales}
\label{subsection:resolution_effects}

\begin{figure}
\epsscale{1.2}
\plotone{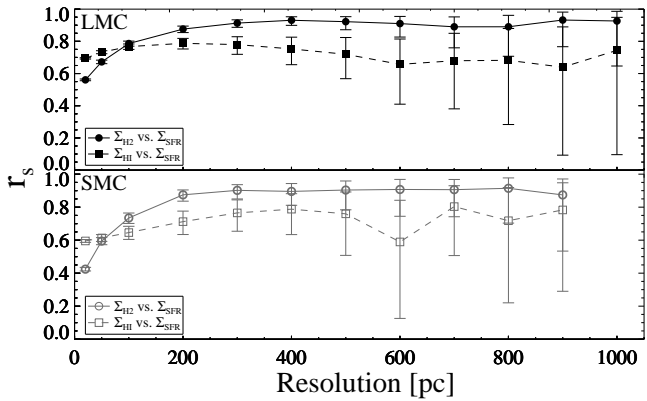}
\caption{ The Spearman rank correlation coefficient ($r_{s}$) as a function of image resolution for the \sightwo\ vs. \sigsfr\ (circles with solid line) and \sighi\ vs. \sigsfr\ (squares with dashed line) distributions. The top plot shows the rank correlations for the LMC and the bottom show those for the SMC. The error bars show the $99.75\%$ confidence interval ($\sim{3\sigma}$) of the measured rank correlation coefficient. The correlation between \hi\ and SFR remains at a constant, high level of $\text{r}_{\text{s}}\sim0.7$ across size scales in part due to the extended nature of both the \hi\ gas and \ha\ emission that dominates the SFR. The correlation between \htwo\ and SFR reaches a maximum value of $\text{r}_{\text{s}}\sim0.9$ at a size scale of 200 pc, which is the expected size scale to average over enough individual star-forming regions to sample a range of evolutionary states.
\label{fig:rank_corr}}
\end{figure}

We use the Spearman's rank correlation coefficient to quantitatively gauge how well the gas correlates with star formation rate at different size scales. Spearman's rank correlation coefficient ($r_{s}$) measures the degree to which two quantities monotonically increase ($r_{s}>0$) or decrease ($r_{s}<0$). We computed the $3\sigma$ confidence intervals using the Fisher $z$-transformation, which is appropriate for bivariate normal distributions. Figure \ref{fig:rank_corr} shows the rank correlation coefficient as a function of resolution for the relationship between star formation rate and molecular gas and atomic gas. 

The change in the rank correlation coefficient with resolution is similar for both the LMC and SMC. As expected for atomic-dominated galaxies, the correlation of \siggas\ vs. \sigsfr\ follows that of \sighi\ and \sigsfr, therefore we only show \sighi\ vs. \sigsfr\ in Figure \ref{fig:rank_corr}. The correlation between \sighi\ vs. \sigsfr\ in both the LMC and SMC is high ($r_{s}\sim{0.6-0.7}$) at the smallest size scale of 20 pc and remains high across all size scales. The correlation of \hi\ with the star formation rate, even at small spatial scales, is due to the extended nature of both components combined with the general trend that regions with more total gas have more star formation and more molecular gas. 

The \sightwo\ vs. \sigsfr\ distribution reaches the maximum correlation coefficient of $r_{s}\sim{0.9}$ at a size scales $\sim{200}$ pc, past which it is better correlated than the relationship with \hi. While the \hi\ is correlated with the star formation rate tracer, we see that molecular gas is best correlated with recent star formation in the LMC and SMC at size scales $\gtrsim{200}$ pc. The 200 pc scale indicates the average size scale where both molecular gas and the star formation rate tracer, \ha, are found together and enough independent star-forming regions at different evolutionary stages (i.e., different ratios of \ha\ to molecular gas) are averaged together. While the correlation peaks at 200 pc, the average molecular gas depletion time decreases until it reaches the integrated value at a size scale of $\sim{500-700}$ pc in the LMC and SMC. The molecular gas and star formation rate tracer have a strong positive correlation, stronger than that with \hi, supporting the physical connection between molecular gas and recent massive star formation. 

\section{Discussion}
\label{section:discussion}

We discuss our findings on the relationship between gas and star formation in the Magellanic Clouds using our new dust-based molecular gas maps. By comparing our results to existing observational studies of mainly massive, high metallicity, molecular-dominated galaxies, simulations, and theoretical models of star formation, we provide insight into the physical mechanisms that drive star formation.

\subsection{\taumoldep\ in the Magellanic Clouds}

\begin{figure}
\epsscale{1.25}
\plotone{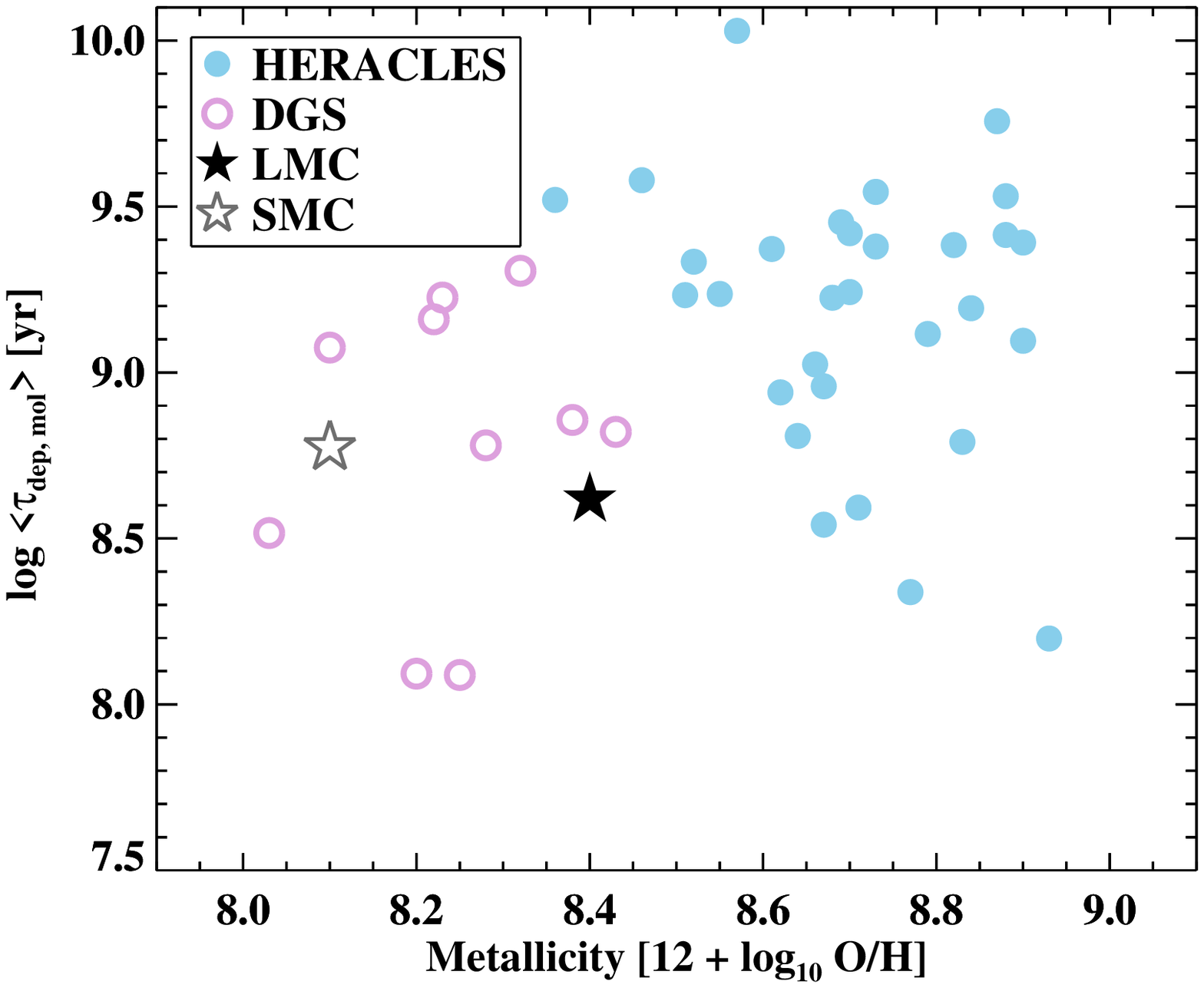}
\caption{The galaxy-averaged molecular gas depletion time ($<\sigmol>/<\sigsfr>$) with metallicity for the HERACLES sample (light blue points), LMC (black filled stars), and SMC (grey open stars). We have taken the average \sigmol\ and \sigsfr\ of the 1 kpc LMC and SMC data, which are comparable measurements to the $\sim{1}$ kpc resolution HERACLES data. We also include the integrated molecular gas depletion times ($M(\htwo)/SFR$) from the Dwarf Galaxy Survey (DGS) using a metallicity dependent CO-to-\htwo\ conversion factor from \citet{cor14}. While there is a large amount of scatter, there does not appear to be any strong trend with metallicity.
\label{fig:dep_metallicity}}
\end{figure}

The range of possible molecular gas depletion times for the LMC and SMC at 1 kpc scales given the systematic uncertainty in our estimate of the molecular gas of $\sim{0.2-1.2}$ Gyr falls below the average $\sim{2}$ Gyr found for nearby normal disk galaxies. This is consistent with the previous work by \citet{bol11} that found $ \taumoldep=1.6$ Gyr at 1 kpc scales in the SMC using similar dust-based molecular gas estimates, with the value being higher due to a higher estimate of the molecular gas. The shorter molecular gas depletion times do not appear to be directly due the lower metallicities as there is no trend in \taumoldep\ with metallicity (see Figure \ref{fig:dep_metallicity}). 

The other remaining environmental factors, besides metallicity, that could affect the ratio of the amount of molecular gas to the amount of current star formation are the lower galaxy masses of the Magellanic Clouds and the interaction between the LMC, SMC, and Milky Way \citep{bes12}. Lower mass galaxies tend to have lower dark matter and stellar densities, making them more susceptible to stochastic bursts of star formation. Both the star formation histories of the SMC and LMC \citep{har04,har09} indicate that there have been recent bursts in star formation in both galaxies. A burst in star formation over a short period of time could lead to a depletion of the molecular gas reservoir combined with higher star formation rates that together can produce low \taumoldep\ values.

The molecular gas depletion time in M33 is $\sim{0.5}$ Gyr when the diffuse \ha\ emission is included \citep{sch10}, which is comparable to our measurements of the Magellanic Clouds. If the diffuse ionized gas is removed from the \ha\ emission, then the molecular gas depletion time increases to $\sim{1}$ Gyr. This highlights the importance of understanding the connection between the diffuse ionized gas and recent massive star formation as it represents a significant fraction of the \ha\ emission and changes \taumoldep. \citet{rah11} found a similar increase in \taumoldep\ by a factor of $\sim{2}$ when the diffuse ionized gas was removed in the disk galaxy NGC 4254. If the diffuse ionized component is excluded in the star formation rate determination, the \taumoldep\ in M33, LMC, and SMC is $\sim{1}$ Gyr.

Like the Magellanic Clouds, M33 is low mass, atomic dominated, has likely interacted with M31 within the past $0.5-2$ Gyr \citep{dav11}. The LMC, SMC, and M33 show evidence for bursts in the star formation history within the last Gyr and the most recent epochs show lower star formation rates, which suggest that the star-forming gas reservoir has been depleted. The observed shorter depletion times appear to be caused by catching these galaxies after a period of higher star formation rate and does not necessarily indicate that these low-mass, low-metallicity galaxies are forming stars differently from normal disk galaxies.

\begin{figure}[t]
\epsscale{1.2}
\plotone{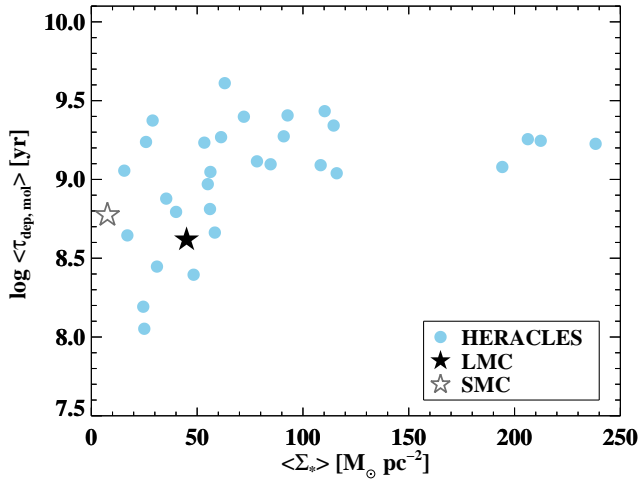}
\caption{Average molecular gas depletion time as a function of the average stellar surface density for the disk-averaged HERACLES sample \citep{ler13a} in blue with the filled and open stars showing the galaxy-averaged data for the LMC and SMC, respectively. The HERACLES sample data used a Galactic CO-to-\htwo\ conversion factor and $\sigstar$ was measured for the HERACLES sample, LMC, and SMC using $I_{3.6\micron}$ from $Spitzer$ and the conversion from \citet{zeb09}. We see that the LMC and SMC points agree with the trend of lower average \taumoldep\ with lower $\sigstar$.
\label{fig:dep_stellarSD}}
\end{figure}

\citet{sai11} also found that for the volume-limited COLD GASS survey, lower stellar mass galaxies ($\sim10^{10}\msol$) had shorter depletion times of $\sim{0.5}$ Gyr. While consistent with the integrated depletion times in the LMC and SMC, the data are not completely comparable since a value for the CO-to-\htwo\ conversion factor has to be assumed and single dish CO observations from the COLD GASS survey will mainly detect the central regions of the galaxies. Saintonge et al. (2011) conjecture that the shorter depletion time is due to the tendency for smaller galaxies to have more ``bursty'' star formation. Similarly, \citet{cor14} suggest that the observed short molecular gas depletion depletion times for their DGS sample of dwarf galaxies are due to recent bursts in star formation. \citet{kau03} found that low redshift galaxies with stellar mass $<3\times10^{10}$ \msol\ in the Sloan Digital Sky Survey (SDSS) have younger stellar populations and that the star formation histories are correlated with the stellar surface density, also indicative of recent bursts in star formation like seen in the LMC, SMC, and M33. In Figure \ref{fig:dep_stellarSD}, we show the average \taumoldep\ time as a function of the average stellar surface density (\sigstar) for the LMC, SMC, and the HERACLES sample of galaxies and see that all of the low molecular depletion times are found at low \sigstar. The fact that low-mass galaxies are more susceptible to stochastic star formation can produce bursts in star formation \citep{hop14} and lead to shorter molecular gas depletion times.

\subsection{Physical Interpretation of the Scatter in \taumoldep} 
\label{subsection:tdep_scatter}

The scatter in the \sigmol-\sigsfr\ relationship, which we quantify in terms of the scatter in $\log\taumoldep$, can be produced by both physical mechanisms and the imperfect nature of the observable tracers of the physical quantities. The previous observational work that focused on the scatter in \taumoldep, or the ``break down'' of the \sigmol-\sigsfr\ relationship, by \citet{sch10}, \citet{ver10}, and \citet{ono10} studied the $\sigmol-\sigsfr$ relationship in M33 over $\gtrsim100$ pc size scales. \citet{sch10} compared \taumoldep\ found for apertures centered on CO peaks to apertures centered on \ha\ peaks for various aperture sizes from $75-1200$ pc and found that the \taumoldep\ values differed for CO  and \ha\ peaks for $\lesssim{300}$ pc size scales. There are a number of possible causes of the difference between the CO and \ha\ molecular gas depletions times: difference in evolutionary stage of the star-forming region, drift of the young stars from their parent cloud, actual variation in \taumoldep, differences in how the observables map to physics quantities, and noise in the maps. \citet{sch10} identify the evolution of individual star-forming regions as the likely cause for the variations.  

At high resolution (scales of $\sim{20-50}$ pc), the star formation and molecular gas are resolved into discrete regions that span a range evolutionary stages \citep[e.g.,][]{kaw09,fuk10} and have different ratios of molecular gas to star formation rate tracers. Averaging over larger size scales samples regions at a range of evolutionary stages resulting in a ``time-averaged'' \taumoldep. The change in the scatter in the molecular gas depletion time ($\sigma$) with resolution informs us about whether the star-forming regions are spatially correlated due to synchronization of star formation by a large-scale process. We see this effect in Figure \ref{fig:mol_sfl}; the scatter in \taumoldep\ decreases as the size scales are increased.

Theoretical studies can be used to explore which mechanism produces the scatter in the \sigmol-\sigsfr\ relationship. We compare our results in the SMC and LMC to the hydrodynamical simulations of galaxies by \citet{fel11} and to the analytical model by \citet{kru14}. Both provide predictions of the amount of scatter in the \sigmol-\sigsfr\ relationship for small size scales (50 pc for predictions from \citealt{kru14}, and 300 pc for the simulations from \citealt{fel11}) and how the scatter changes with size scale. 

The simulations by \citet{fel11} show that the time-averaging of the star formation rate (or, our inability to measure the instantaneous star formation rate) combined with \sigmol\ estimates that are instantaneous alone can generate most of the scatter observed in the \sigsfr-\sightwo\ relation. If we possessed a perfect, instantaneous tracer of the star formation rate, then we would expect to see high star formation rates while there is still a large amount of molecular gas. As the molecular gas is depleted and destroyed by the previous episode of star formation, both the molecular gas and star formation rate would decrease. Instead, we observe the tracers of the star formation rate (namely \ha) peak when the molecular gas is partially or mostly dissipated because the tracers show the average star formation rate over up to $\sim10$ Myr \citep{ken12}. The time evolution of star-forming regions alone does not cause the observed offset between the observed star formation rate and molecular gas, rather the time-averaging of the star formation rate combined with the time evolution of star-forming regions produces different ratios of molecular gas to star formation rate and scatter in \taumoldep. \citet{hon15} shows evidence of this effect in the N66 region of the SMC where the star formation rate from \ha\ disagrees with that from pre-main sequence stars at small ($\sim{6-150}$ pc) size scales. 

\citet{kru14} quantify how the scatter in the $\siggas-\sigsfr$ relationship should change with size scale due to the incomplete statistical sampling of independent star-forming regions, including the effect of the different timescales associated with the gas and star formation tracers discussed by \citet{fel11}, and add the additional scatter associated with incomplete sampling of star formation rate tracers from the initial stellar mass function (IMF), and the spatial drift between stars between gas and stars. The model requires having an estimate of the lifetime of GMCs ($t_{\textnormal{gas}}$), the time scale for the star formation rate tracer ($t_{\textnormal{stars}}$), the time where both the gas and star formation rate tracer overlap ($t_{\textnormal{over}}$), the typical separation between independent star-forming regions ($\lambda$), the flux ratio between peaks in the overlap phase and in isolation for the gas and star formation ($\beta_{1}$, $\beta_{2}$), the scatter due to the time evolution of gas and star formation flux ($\sigma_{\textnormal{evol,1g}}$, $\sigma_{\textnormal{evol,1s}}$), the scatter due to the mass spectrum ($\sigma_{\textnormal{MF}}$), and the observational error ($\sigma_{\textnormal{obs}}$). The predictions for the scale dependence of the scatter in the gas depletion time from \citet{kru14} agrees with the predictions by \citet{fel11} at size scales $>300$ pc where the two are directly comparable. \citet{kru14} find that the scatter varies from $\sim{0.9}$ dex at 50 pc scales to $\sim{0.2}$ dex at 1 kpc. The trend in the prediction of scatter (valid for their fiducial parameter values for disks and dwarfs) is consistent with the observation from the HERACLES galaxies \citep{ler13a} and M33 \citep{sch10}. 

For the LMC and SMC predictions of the scatter from \citet{kru14}, we use estimates of the input parameters based on observational data when possible. For both the LMC and SMC, we set $t_{\textnormal{stars}} = 6$ Myr based on the lifetime of \ha, make the assumption that $\beta_{1}$, $\beta_{2}=1$, and set $\lambda=150$ pc, the typical Toomre length (for $\Sigma_{\textnormal{gas}}\sim{10}~\smpc$ and $\Omega\sim{0.03}$ Myr$^{-1}$). In the LMC, we use the results of \citet{kaw09} to set $t_{\textnormal{gas}} = 26$ Myr and $\sigma_{\textnormal{MF}} = 0.4$ dex (the mean logarithmic scatter of the Class I GMC mass). For the upper limit in the LMC, we take $t_{\textnormal{over}}=0$ Myr and $\sigma_{\textnormal{evol,1g}}$, $\sigma_{\textnormal{evol,1s}}=0.3$ dex, based on a linear time evolution to or from zero. For the lower limit in the LMC, we adopt $t_{\textnormal{over}}=3$ Myr (the supernova timescale) and $\sigma_{\textnormal{evol,1g}}$, $\sigma_{\textnormal{evol,1s}}=0.15$ dex; while it can vary from $0-0.3$ dex, half the amount of scatter as linear evolution is a reasonable lower limit since the parameters must be $>0$ due to the existence of molecular clouds without massive stars and \hii\ regions without molecular clouds. We note that \citet{kru14} assume that the galactic star formation rate is roughly constant over the entire lifetime of the GMCs ($\sim{30}$ Myr), which stands in contrast to the multiple bursts over the past $\sim{50}$ Myr identified in the star formation history of the both LMC \citep{har09} and SMC \citep{har04}.

In the LMC, where the morphology is more clearly a disk and the metallicity is not much lower than Solar, we observe scatter at the level of $\sim{0.45}$ dex at $\sim100$ pc and $\sim{0.18}$ dex at $\sim$ kpc scales. The \citet{fel11} simulations show that the behavior of the scatter in $\log{\taumoldep}$ with averaging size scale from $\sim{100-1000}$ pc for Solar metallicity and radiation field are remarkable similar to the observations for the LMC. The simulations from \citet{fel11} predict scatter of $\sim{0.4-0.6}$ dex at $\sim{100}$ pc scales and $\sim{0.1-0.3}$ dex at $\sim$ kpc scales for their fiducial solar metallicity simulations (across the range of their parameter exploration). The \citet{kru14} model produces a range in the predicted scatter in $\log{\taumoldep}$ in the LMC of $0.46-0.51$ dex at 100 pc scales and $0.19-0.23$ dex at 1 kpc scales, which are comparable to the results from the \citet{fel11} and close to the observed values for the LMC (see Figure \ref{fig:tdep_scatter_res}). The dominant source of scatter at large ($>100$ pc) size scales for the lower limit predictions (closest to the observations) from the \citet{kru14} model comes from the Poisson statistics of the number of times each evolutionary phase of star formation is sampled, which is determined primarily by the timescale of the star formation rate tracer, the lifetime of GMCs, and the separation between star-forming regions. The similarity between our observations and both the \citet{fel11} simulation and \citet{kru14} model at large ($>100$ pc) size scales, where both are comparable and individual star-forming regions are unresolved, supports the interpretation that the scatter in the $\sigmol-\sigsfr$ relationship can be largely attributed to star formation rate tracers that time-average the ``true'' or instantaneous star formation rate.

\subsubsection{Scatter in \taumoldep as a Function of Size Scale}

\begin{figure}[t]
\epsscale{1.25}
\plotone{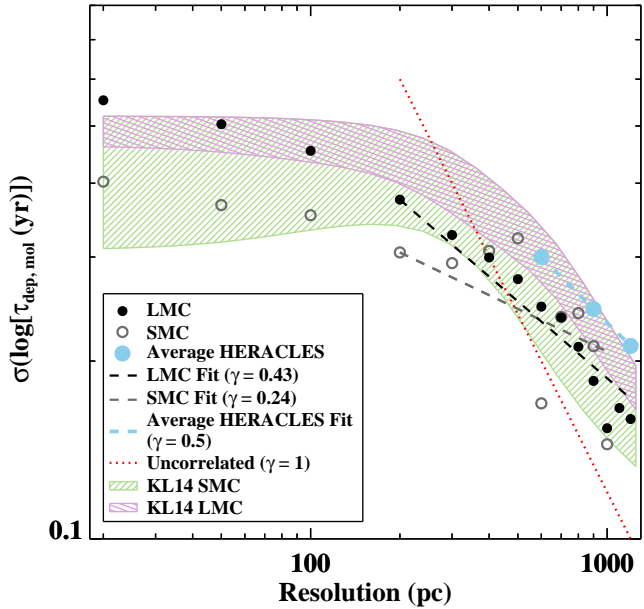}
\caption{We fit power laws (dashed lines) to the change in scatter in the depletion time as a function of scale for the LMC (black filled circles) and SMC (grey filled circles), and find that the power law exponent ($\gamma$) is low indicating correlation of star formation throughout the galaxies likely due to synchronization by a large-scale process. The thick light blue line shows the line for $\gamma=0.5$, the average fit to the 9 HERACLES galaxies that had high enough resolution ($r\sim400$ pc) and the expected scaling for a disk galaxies from simulations by \citet{fel11}. The purple and green hashed lines show the estimates of the scatter due to independent star-forming regions from the \citet{kru14} model (KL14) with upper and lower limits for the LMC and SMC, respectively. For comparison, the red dashed line shows how the data would behave if there was no spatial correlation between the star formation and molecular gas at large spatial scales. 
\label{fig:tdep_scatter_res}}
\end{figure}

As a means to quantify the behavior of scatter with different size scales, \citet{fel11} fit a power law to the relationship between size scale and the scatter in $\log{\taumoldep}$.  \citet{ler13a} used a subset of nearby HERACLES galaxies to study the scatter in \taumoldep\ at linear resolutions of $0.6-2.4$ kpc, which we can compare to our results in the Magellanic Clouds spanning linear resolutions of $0.02-1$ kpc. Following  \citealt{fel11} and \citet{ler13a}, we quantify the scale-dependence of the scatter in \taumoldep\ in the LMC and SMC by 
\[ \sigma(l) = \sigma_{100} \left( \frac{l}{100\textnormal{ pc}} \right)^{-\gamma} \]
where $l$ is the spatial resolution, $\sigma_{100}$ is the scatter in $\log{(\taumoldep)}$ at 100 pc resolution, and the power-law index $\gamma$ measures the rate that the scatter changes with resolution ($\gamma=1$ for uncorrelated star formation in a disk)\footnote{We use $\gamma$ instead of $\beta$ (used in \citet{ler13a}) as the variable representing the exponent to avoid confusion with the dust emissivity index $\beta$.}. We fit only resolutions greater than 100 pc, since below that resolution $\log{(\taumoldep)}$ will be biased by negative and zero values. Figure \ref{fig:tdep_scatter_res} shows how the scatter in \taumoldep\ changes with resolution, including the best fit power-law functions with $\gamma=0.43$ for the LMC and $\gamma=0.24$ for in the SMC. \citet{ler13a} find a best fit $\gamma$ for the scatter in $\log(\taumoldep)$ in the range of 0-0.8 with an average of $\gamma=0.5$ (shown by the thick dashed line in Figure \ref{fig:tdep_scatter_res}).

If a galaxy has a fixed \taumoldep\ and star formation proceeds randomly and independently in separate regions within the resolution element, then behaves like Poisson noise and $\sigma \propto{\sqrt{N^{-1}}}$, where $N$ is the number of star forming regions. For a region of size $l$, $N\propto{l^{2}}$ so that $\sigma \propto{l^{-1}}$ or a power-law scaling of $\gamma={1}$. Both \citealt{fel11} and \citet{kru14} find that the scatter in $\log{(\taumoldep)}$ scales with a rough power-law scaling with an index of $\gamma=0.5$ at larger ($\gtrsim200$ pc) scales. \citet{fel11} expect this shallow scaling as a result of star formation occuring in a 2D disk galaxy. However, \citet{kru14} finds similar shallower slopes with uncorrelated, independent star-forming region due to the contribution to the scatter from the time evolution of the star-forming regions and from the underlying distribution of GMC masses. The model from \citet{kru14} shows that the scatter due to Poisson noise dominates at large size scales that sample multiple star-forming regions ($>100$ pc). At small scales, the Poisson noise disappears due to the fact only one star-forming region will be sampled and the scatter from the time evolution of the star-forming regions and from the underlying distribution of GMC masses drives the variation in ratio of star formation rate to molecular gas and \taumoldep. 

Figure \ref{fig:tdep_scatter_res} shows the data for the LMC and SMC, the HERACLES galaxies, and the corresponding \citet{kru14} model predictions. For the SMC model predictions, we set the upper limit to the upper limit values for the LMC and the lower limit the same as for the LMC but with $t_{\textnormal{gas}}=10$ Myr (approximate free-fall time of a GMC) and $\sigma_{\textnormal{MF}} = 0.2$ dex ($60\%$ of the scatter in GMC masses in the LMC). We see general agreement between the trend in the observed relationship and the model predictions, however, the LMC data fall below the predicted lower limit (see Section \ref{subsection:tdep_scatter}) at large size scales ($>200$ pc). The most uncertain parameter in \citet{kru14}, due to lack of observational constraints, is the scatter due to the time evolution of the gas flux and star formation rate flux ($\sigma_{\textnormal{evol,1g}}$, $\sigma_{\textnormal{evol,1s}}$). Decreasing $\sigma_{\textnormal{evol}}$ further from 0.15 dex to 0.1 dex brings the model predictions much closer to the LMC observations at scales $>100$ pc.

If we apply the interpretation of \citet{fel11}, the shallower decline in the amount of scatter with increasing averaging scale seen in the LMC (and SMC) could be caused by increased spatial correlation between individual star-forming regions. Correlation of star-forming regions, both spatially and temporally, would cause individual star-forming regions to be at similar evolutionary phases throughout large parts of the galaxy and could explain the need for a lower amount of scatter from the time evolution for the gas flux and star formation rate. Large-scale spatial correlation in star formation requires a physical mechanism to synchronize star formation, such as bursts of star formation throughout large parts of the galaxies driven by tidal interactions or ram pressure. The star formation histories of both the LMC and SMC indicated that there have been recent bursts of star formation throughout large parts of the galaxies, likely due to interaction between the galaxies and the Milky Way, and is possibly driving the shorter molecular gas depletion times. The lower amount of scatter in $\log{\taumoldep}$ at larger size scales observed in the LMC (and more tenuously in the SMC) could also be due to large-scale synchronization of star formation. 

\begin{figure*}[tp]
\epsscale{1.1}
\plottwo{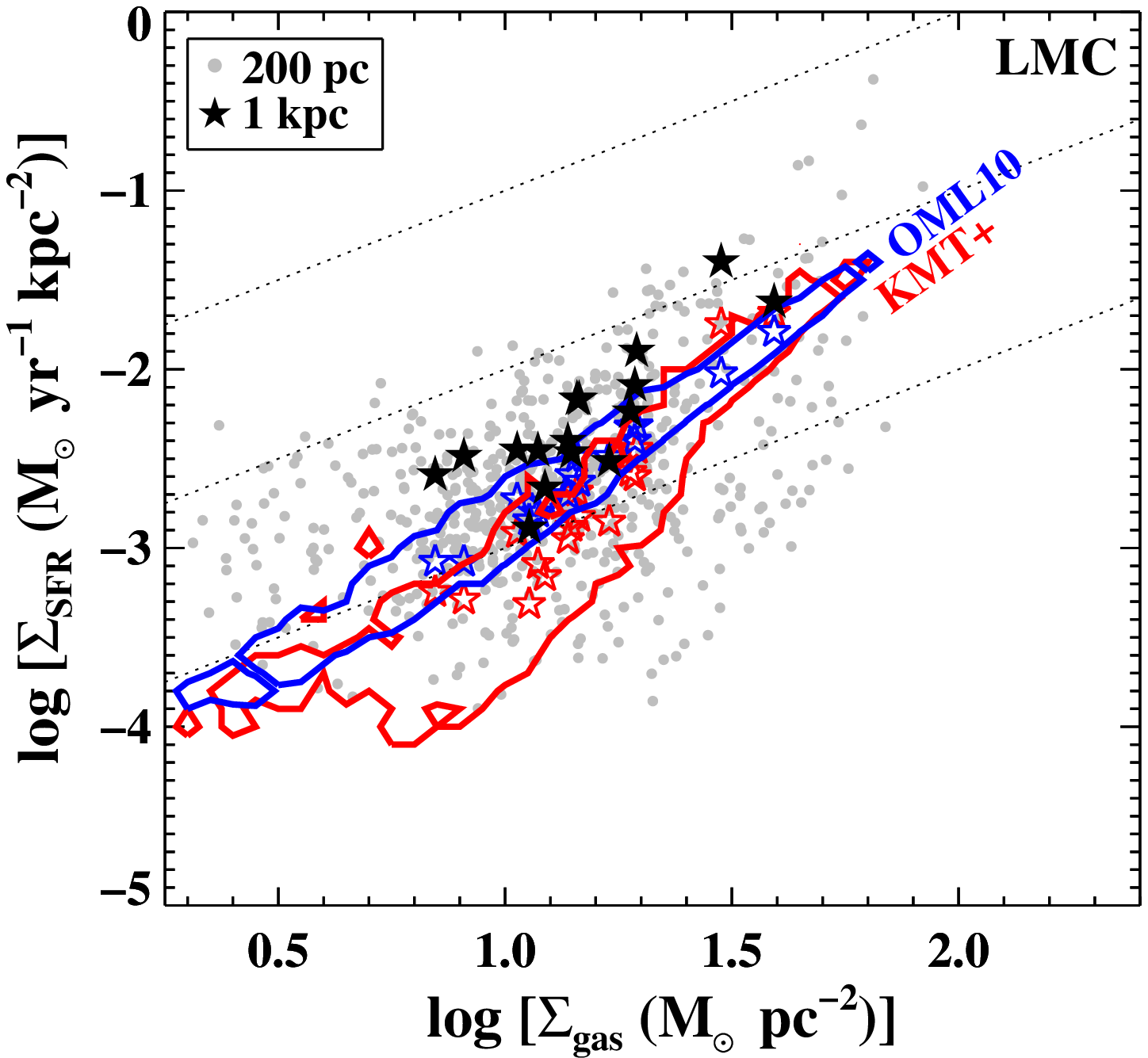}{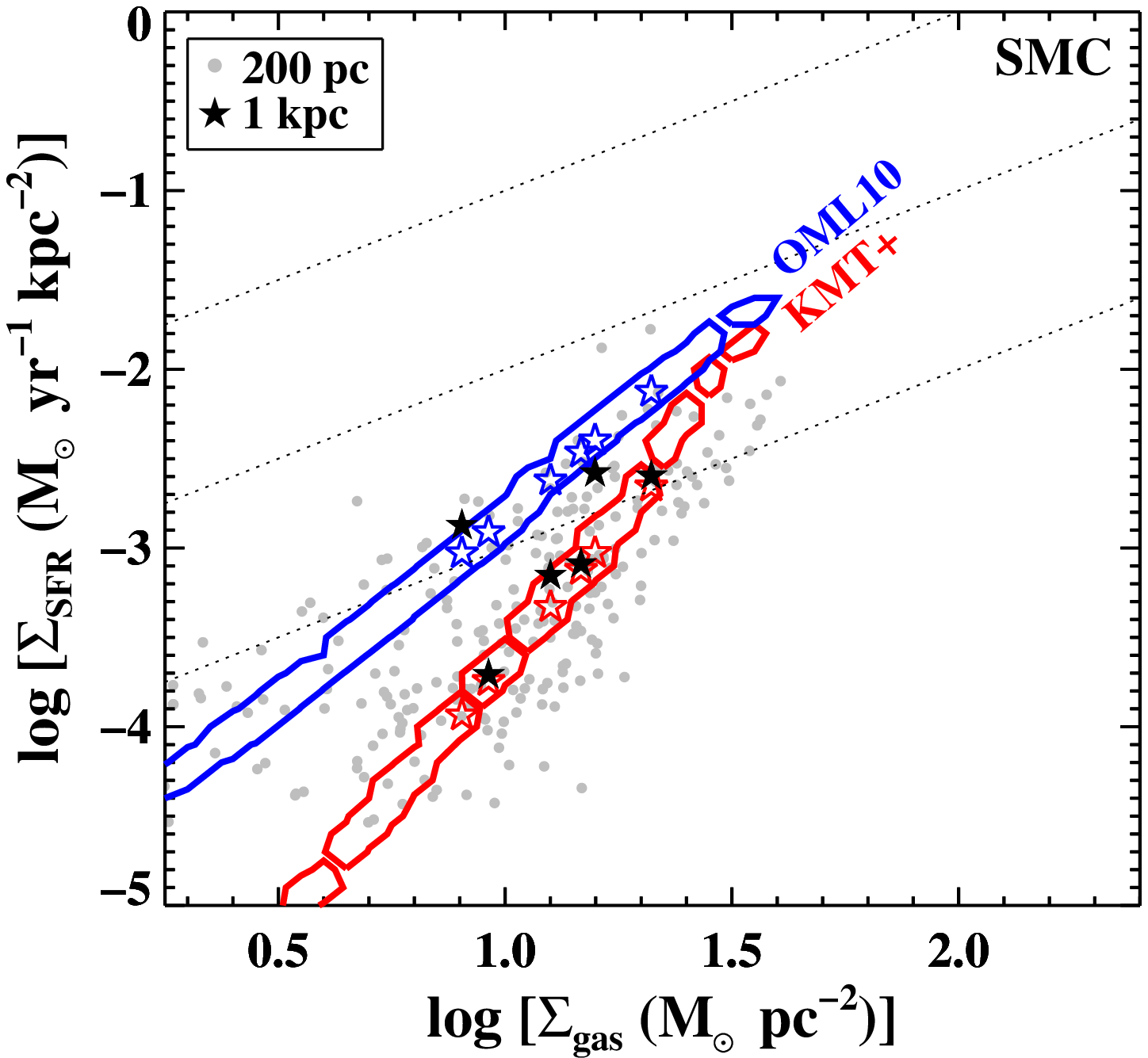}
\caption{Star formation rate predictions from the OML10 and KMT+ models. The grey filled circles show the data at $r=200$ pc, and the filled stars show the $r=1$ kpc data. The diagonal dashed lines indicate constant total gas depletion times (from bottom to top: 10 Gyr, 1 Gyr, and 0.1 Gyr). The contours show the full extent of the distribution of points for the model predictions at $r=200$ pc resolution and the open stars show the $r=1$ kpc predictions. The KMT+ and OML10 predictions are shown in red and blue, respectively, for the appropriate metallicities for each galaxy and for $t_{\textnormal{SF,GBC}}=0.5$ Gyr for the OML10 model. Both models predict the trend in the data, but do not capture the full extent of the scatter observed. 
\label{fig:model_predictions}}
\end{figure*}

\subsection{Comparison to Star Formation Model Predictions}

The Magellanic Clouds provide ideal laboratories to test models of star formation given their low metallicity. While higher metallicity, the geometry of the LMC is better understood than the more irregular SMC. There are few nearby low-mass, low-metallicity systems and measuring their molecular gas content is challenging as they are often weakly emitting in CO and, when CO is observed, it is unclear what CO-to-\htwo\ conversion factor should be applied. The dust-based molecular gas estimates for the LMC and SMC allow us to test metallicity-dependent models of star formation at high resolution. We examine the model predictions from \citet[hereafter OML10]{ost10} and \citet[hereafter KMT+]{kru13}, a recent update of the \citet{kru09} model modified for atomic-dominated regions. Both models take the total gas surface density (\siggas) and metallicity ($Z'$) as input parameters and predict the fraction of molecular gas, and from that the star formation rate. 

The OML10 model determines the star formation rate based on a balance between vertical gravity in the disk and the pressure of the diffuse ISM, which is controlled by star formation feedback. OML10 relates thermal pressure to \sigsfr, whereas \citet{ost11} relates turbulence to \sigsfr.  While the star-forming gas in the OML10 model is not strictly molecular gas, but rather bound clouds, we identify our estimate of \sightwo\ with the model parameter $\Sigma_{\text{gbc}}$, the surface density of gas in gravitationally bound clouds (note that in our methodology both H$_2$ and any optically thick \hi\ are effectively indistinguishable). This ignores the fact that in very dense regions a significant fraction of the molecular gas could be not self-gravitating, a concern that is probably important in starburst environments but unlikely to matter in the Magellanic Clouds. The reverse concern, that \hi\ may make a significant contribution to the cloud bounding mass, is likely a more significant consideration in these sources, although its magnitude is difficult to evaluate. The KMT+ model is based on the assumption that the fraction of molecular gas is mainly determined by the balance between the dissociating UV radiation field and the shielding of the gas. The KMT+ model adds to \citet{kru09} the condition that in a region with low star formation rate, hence low UV field, the threshold density of the cold neutral medium is no longer set by two-phase equilibrium between the cold and warm neutral medium, but rather by hydrostatic equilibrium.


Both the KMT+ and OML10 model use the mid-plane pressure, which requires an estimate of the density of stars and dark matter in the disk to determine the gravitational pressure. We estimate the stellar surface density by applying the mass-to-light conversion from \citet{ler08} to the 3.6 \micron\ $Spitzer$ SAGE images of the LMC and SMC. The $\Sigma_{*}$ is then converted to volume density by assuming a stellar disk thickness of 600 pc for the LMC \citep{van02} and 2 kpc for the SMC (following \citealt{bol11}). For the LMC, we use the dark matter density profile from \citet{alv00}, for the SMC we use the profile from \citet{bek09} to estimate the dark matter density as a function of radius from the centers of the galaxies. We find that the combined stellar and dark matter densities have ranges of $0.6-0.1$ \msol\ pc$^{-3}$, with the higher values concentrated in the stellar bar and $\sim{10\%}$ dark matter contribution in the LMC, and $0.006-0.1$ \msol\ pc$^{-3}$ with $\sim{20\%}$ dark matter contribution in the SMC. 

We adopt most of the fiducial model parameter values as described in OML10 and KMT+. The exception is the depletion time in gravitationally bound clouds, $t_{\textnormal{SF,gbc}}$, for which OML10 uses 2 Gyr based on the average observed value in nearby galaxies. This value was also applied in the SMC results in \citet{bol11} since the observed \taumoldep\ was not much lower than 2 Gyr. In this study we find a wider range of depletion timescales, and a measurable change in \taumoldep\ as a function of spatial scale (Fig. \ref{fig:dep_res}). It is important to note that $t_{\textnormal{SF,gbc}}$ is an input parameter of the OML10 model, obtained from observations rather than theory, and its main impact is to change the relation between $\Sigma_{\text{gbc}}$ and $\Sigma_{\text{SFR}}$ since $\Sigma_{\text{gbc}}=t_{\textnormal{SF,gbc}}\Sigma_{\text{SFR}}$ 

The self-regulation in the model operates to make $\Sigma_{\text{SFR}}$ insensitive to the choice of $t_{\textnormal{SF,gbc}}$ over a wide range of total gas surface densities. The results we show in Fig. \ref{fig:model_predictions} are computed for $t_{\textnormal{SF,gbc}}=0.5$~Gyr, which corresponds approximately to the value of \taumoldep\ we observe at large spatial scales. The main effect of changing $t_{\textnormal{SF,gbc}}$ from 0.5 to 2 Gyr is to slightly lower the predicted star formation rate, particularly at high surface densities ($\siggas \gtrsim{50}$ \smpc).  The robustness of $\Sigma_{\text{SFR}}$ to the choice of $t_{\textnormal{SF,gbc}}$ in turn means that $\Sigma_{\text{gbc}}$ depends significantly on the value of the depletion timescale. Since we identify $\Sigma_{\text{gbc}}$ with \sightwo, the consequence is that comparison of our measurements of \sightwo\ with the model predictions are extremely dependent on the assumed \taumoldep, and on the constancy of \taumoldep\ with \siggas. In other words, in the context of the model they are very uncertain. Adopting the approximate value observed at large spatial scales, $\taumoldep\approx0.5$~Gyr, results in a predicted $\Sigma_{\text{gbc}}$ very similar to the observed \sightwo.

Figure \ref{fig:model_predictions} shows the model predictions for the \sigsfr\ for the LMC and SMC. Since both models require averaging over the different gas phases, we only compare the predictions to the data at $r\sim200$ pc and $r\sim1$ kpc. While the models can predict the molecular-to-atomic ratio (or, diffuse to gravitationally bound for OML10), the models self-regulate in diffuse gas, which is the dominant regime in the LMC and SMC, and predict similar star formation rates. Both KMT+ and OML10 models predict the general trend observed in the relationship between \siggas\ and \sigsfr .


The models do not predict the amount of scatter seen in the higher resolution 200 pc data. This is not surprising since both OML10 and KMT+ predict a time-averaged star formation rate and do not recover variations in the star formation rate based on the details on the star formation rate tracer combined with differences in the evolutionary stages of individual star-forming regions. Assuming that the physical interpretation from \citet{fel11} is correct, which is supported by the predictions of the amount of scatter in \taumoldep\ at $\sim{100}$ pc and 1 kpc scales matching our observations, then an important, possibly the dominant, source of scatter is the time-averaging of the star formation rate (over as little as 10 Myr, \citealt{ken12}) inherent in using \ha\ and 24 \micron\ as star formation rate tracers. We think that it is likely that the under-prediction of the amount of scatter in the OML10/KMT+ model predictions at 200 pc is due to the fact that the star formation models do not include the time-averaging effect of the star formation rate tracer (\ha). The scatter present in the OML10/KMT+ model predictions come only from the spatial variation in the stellar and dark matter density, which affects the pressure and therefore the predicted amount of star-forming gas. When averaging over larger ($\sim$ kpc) scales, the difference between the scatter in the data and the scatter in the predictions decreases and the two are comparable. The model predictions are most appropriate at large scales where many independent star-forming regions are averaged over to account for the fact that star formation is treated as a time-averaged process.

The main differences between the star formation rates from OML10 vs. KMT+ appear at high \siggas\ and low metallicity, where the predictions diverge. All of the KMT+ predictions, independent of metallicity, converge at high \siggas\ because once the ISM transitions to \htwo-rich, which happens past a column density based on the amount of shielding, then the star formation will not behave any differently from high metallicity galaxies. The OML10 model will tend to continue to predict lower \sigsfr\ at high \siggas\ because the lower metallicity increases the thermal pressure and reduces the star formation at all surface densities. At the same metallicity and at high \siggas, OML10 will predict lower \sigsfr\ than KMT+.


\section{Summary and Conclusions}
\label{section:conclusions}

We create \htwo\ maps for the LMC and SMC by using dust emission from HERITAGE $Herschel$ images as a tracer of the total column density of gas and subtracting off the atomic component, which avoids the known biases of using CO as a tracer of the molecular gas at lower metallicity. Our dust-based methodology has the potential to include optically thick and/or very cold \hi, but we see no evidence of this and assume that all of the gas in our maps is molecular. We find total molecular gas masses of $M_{\mathrm{LMC}}^{\mathrm{mol}} = 6.3 ^{+6.3}_{-3.2} \times 10^{7}$ \msol\ and $M_{\mathrm{SMC}}^{\mathrm{mol}} \sim 1.3^{+1.3}_{-0.65} \times 10^{7}$ \msol\ including the estimated systematic uncertainty. The structure of the molecular gas maps show good agreement with the structure of the MAGMA \co\ map in the LMC, with the main difference that more extended \htwo\ is seen using our dust-based method. 

Using our \htwo\ maps we study the relationship between gas and star formation without relying on a conversion factor to translate CO emission to the total amount of molecular gas. The high resolution data allow us to study the relationship over 20 to 1000 pc scales. Our main conclusions are as follows:
\begin{enumerate}
\item Combining the new molecular gas estimates with the star formation rate from \ha\ and 24 \micron, we find molecular gas depletion times at 1 kpc scales of 0.4 Gyr for the LMC and 0.6 Gyr for the SMC (Figure \ref{fig:dep_res}). These molecular gas depletion times are shorter than the the average found for normal, nearby star-forming galaxies (Figure \ref{fig:mol_sfl}), but are within the scatter found in the STING and HERACLES samples \citep[;Figure \ref{fig:leroy13_sfl_h2}]{rah12,ler13a}. We show that when we include our dust-based molecular gas depletion time measurements with those using CO from the HERACLES sample (Figure \ref{fig:dep_metallicity}) we see no trend with metallicity, which suggests the possible trends seen by \citep{rah12,ler13a} could be due to the affect of metallicity on the CO-to-\htwo\ conversion factor. The shorter molecular gas depletion times in the LMC and SMC are similar to that observed in M33 \citep{sch10} and may be associated with the recent bursts in their star formation history.

\item We measure the rank correlation coefficient of the relationships between \sighi\ and \sigsfr\ and \sightwo\ and \sigsfr\ from 20 to 1000 pc size scales (\ref{fig:rank_corr}). The correlation between \sighi\ and \sigsfr\ is scale independent while the correlation between \sightwo\ and \sigsfr\ increases steadily until flattening out at scales of $\sim{200}$ pc and larger and on those scales is better correlated than \sighi\ and \sigsfr. 

\item We measure the scatter in the molecular gas depletion time as a function of size scale (Figure \ref{fig:tdep_scatter_res}). We have compared the observed scatter in the molecular gas depletion time to the predictions from the simulations by \citet{fel11} and the model by \citet{kru14}. We find that both can produce the behavior of the scatter with size scale, which suggests that scatter in the $\sigsfr-\sightwo$ relation may be driven largely by the time-averaging effect of the star formation rate tracer combined with instantaneous measurements of the molecular gas at large scales ($>100$ pc). From comparison with \citet{fel11} and \citet{kru14}, we see possible evidence of synchronization of star formation in how the amount of scatter changes with size scale in the LMC (and potentially the SMC), perhaps due to star formation on large-scales caused by interactions.

\item We have compared the observed \sigsfr\ to the predictions from OML10 and KMT+ star formation models (Figure \ref{fig:model_predictions}) and find wide agreement, indicating that the inclusion of a diffuse neutral medium is important for predicting the star formation rate in atomic-dominated systems like the Magellanic Clouds. Neither model captures the full extent of the scatter seen in the data at 200 pc scales, which we attribute to the time-averaging effect of the star formation rate tracer (as referred to in our previous conclusion).

\end{enumerate}

\acknowledgments

We thank Mark Krumholz for useful discussions and for use of his Python code implementation of the KMT+ model and Diederik Kruijssen for helpful comments and assistance using their code from \citet{kru14}. We thank Eve Ostriker for very thoughtful comments. A.D.B. and K.E.J. wish to acknowledge partial support from grants NSF-AST0955836 (CAREER), NSF-AST141241, NASA-JPL 1372988, 1483968, and 1454733. We acknowledge financial support from the NASA Herschel Science Center, JPL contracts \#1381522, \#1381650, and \#1350371. We thank the contributions and support from the European Space Agency (ESA), the PACS and SPIRE teams, the Herschel Science Center (esp. L. Conversi) and the NASA Herschel Science Center (esp. A. Barbar and R.Paladini) and the PACS and SPIRE instrument control centers (esp. George Bendo), without which none of this work would be possible. M. Meixner is grateful for support from NSF grant 1312902.



{\it Facilities:} \facility{Herschel}, \facility{Spitzer}.


 
\clearpage

\appendix

\section{Modeling Thermal Dust Emission}
\label{appendix:modeling_dust}

The dust emission modeling done in this work used a modified blackbody with a fixed emissivity index, $\beta$, to fit the dust temperature $T_{d}$ pixel-by-pixel to the 100, 160, 250, and 350 \micron\ HERITAGE  images. Excluding the 500 \micron\ image avoids the issue of possible ``excess'' dust emission at $\lambda>400~\micron$ observed in the SMC and LMC \citep{pla11, gor14}. From the fitted $T_{d}$, we calculate, $\tau_{160}$ for all $3\sigma$ fits to $T_{d}$ using
\[ \tau_{160}=\frac{S_{160}\textnormal{[MJy/sr]}}{B_{\nu}(T_{dust}, 160 \micron)}\, . \]
All images are convolved to the resolution of the 350 \micron\ data, the lowest resolution (25\arcsec), using the kernels from \citet{ani11}. We fix $\beta$ in the model to avoid the degeneracy between $T_{d}$ and $\beta$, which can occur when the correlated errors between the $Herschel$ bands are not taken into account. Fixing $\beta$ also follows the previous work in the SMC \citep{ler09, bol11}. We adopt $\beta=1.8$ for our fiducial molecular gas map because that is the approximate average value of $\beta_{1}$ found in the broken emissivity modified blackbody (BEMBB) modeling by \citet{gor14} and similar to $\beta=1.7$ found for M33 using the $Planck$ data (F. Israel, private communication). We also create maps using $\beta=1.5$ and $\beta=2.0$ to see how that affects the \htwo\ estimate since $\beta\sim1-2$ for carbonaceous grains \citep{jag98} and $\beta\sim2$ for silicate grains \citep{cou11}.

The second map uses the BEMBB dust emission modeling results from \citet{gor14}, which uses the same $Herschel$ data, but includes the 500 \micron\ image and accounts for correlated uncertainty between the different bands. All images are convolved to the resolution of the 500 \micron\ data (35\arcsec), and thus lower resolution than our first method of dust modeling. The implementation of the correlated uncertainties in \citet{gor14} eliminates the degeneracy between $T_{d}$ and $\beta$, allowing both to be fit by the models. \citet{gor14} fit three different modified blackbody models to the data: a simple modified blackbody, one that allows two temperatures, and one with broken emissivity index (fits two $\beta$ values and the break wavelength). We use the surface mass density of dust (\sigdust) from the broken emissivity model (with $0.8<\beta_{1}<2.5$) because it produces the smallest residuals and the gas-to-dust ratio falls within the range allowed by elemental abundances. To be comparable to the dust modeling done in this work, we convert \sigdust\ map to $\tau_{160}$:
\[ \tau_{160} = \kappa_{\textnormal{eff, }160}\sigdust \]
where $\kappa_{\textnormal{eff, }160}=11.6$ [cm$^{2}$ g$^{-1}$], which \citet{gor14} finds by calibrating the broken emissivity model to reproduce the diffuse Milky Way SED \citep{com11} with a  gas-to-dust ratio of 150, based on the depletion measurements from \citet{jen09}.

\section{Offsets in \hi\ vs. Dust}
\label{appendix:hi_offset}

The \htwo\ mapping method assumes that the gas-to-dust ratio in the diffuse, atomic ISM is the same as in the dense, molecular regions. As part of the mapping, we investigated the global and regional relationship between \Nhi\ and $\tau_{160}$. The global relationship between \Nhi\ to $\tau_{160}$ is primarily defined by one linear relationship (equivalent to a single dust-to-gas ratio) with a large amount of scatter (see Figure \ref{fig:lmc_Nhi_tau}). We split the galaxies into quadrants to fit the offset in \hi . For the LMC, we split the Southeast quadrant into 16 smaller regions due to the complexity of this part of the galaxy: the Molecular Ridge and an \hi\ streamer that extends to become part of the Magellanic Bridge. Figure \ref{fig:lmc_regions} shows the fitted offsets in \sighi\ in the LMC, with the offsets typically being $\sim5\smpc$. 

By splitting up the LMC into four equal quadrants and looking at the regional relationships between \Nhi\ and $\tau_{160}$, we found that the offset distribution is coming from the SW quadrant. We further checked for smaller regional variation within the different molecular gas complexes and found the majority of offset points to be coming from the Molecular Ridge. Possible explanations for an offset at higher \Nhi\ compared to the dust include: issues with the background subtraction in the $Herschel$ data or a constant layer of \hi\ gas with little to no dust along the line of sight. We note that variations in the gas-to-dust ratio would only change the slope of the distribution, and not just the offset. An issue with background subtraction seems unlikely since the excess offset appears to be correlated with a physical complex and there is no obvious gradient across the quadrant. A layer of low-dust or dust-free \hi\ is possible since there is an \hi\ streamer extending out of the galaxy in this area that becomes part of the Magellanic Bridge. It is possible the stripped gas could have little to no dust. 

\vspace{\baselineskip}

\noindent \underline{Steps to Finding \hi\ Offset}
\begin{enumerate}
	\item Mask all points that likely have molecular gas given the CO map (all regions within 2\arcmin\ of bright CO emission ($I_{\text{CO}}>3\sigma$ detections)
	\item First fit a linear equation to the binned medians of the diffuse gas (\Nhi $< 3 \times10^{21}$ cm$^{-2}$); the slope represents the effective GDR (\dGDR) and offset gives the \hi\ offset. 
	\item Use the fitted \dGDR\ to estimate the total gas using the dust map ($\siggas = \dGDR\sigdust$), subtract \sighi\ to get first iteration estimate of \htwo\ ($\sigmol = (\dGDR\sigdust) - \sighi$).
	\item Mask all points near bright CO and any points that have estimated $\sigmol > 0.5\sighi$.
	\item Refit a linear equation to the binned medians of the diffuse gas with the new mask.
	\item Remove the \sighi\ offset from the second iteration of fitting from the \hi\ map and use this subtracted \hi\ map in the rest of the analysis.
\end{enumerate}

\begin{figure*}[tp]
\epsscale{1}
\plottwo{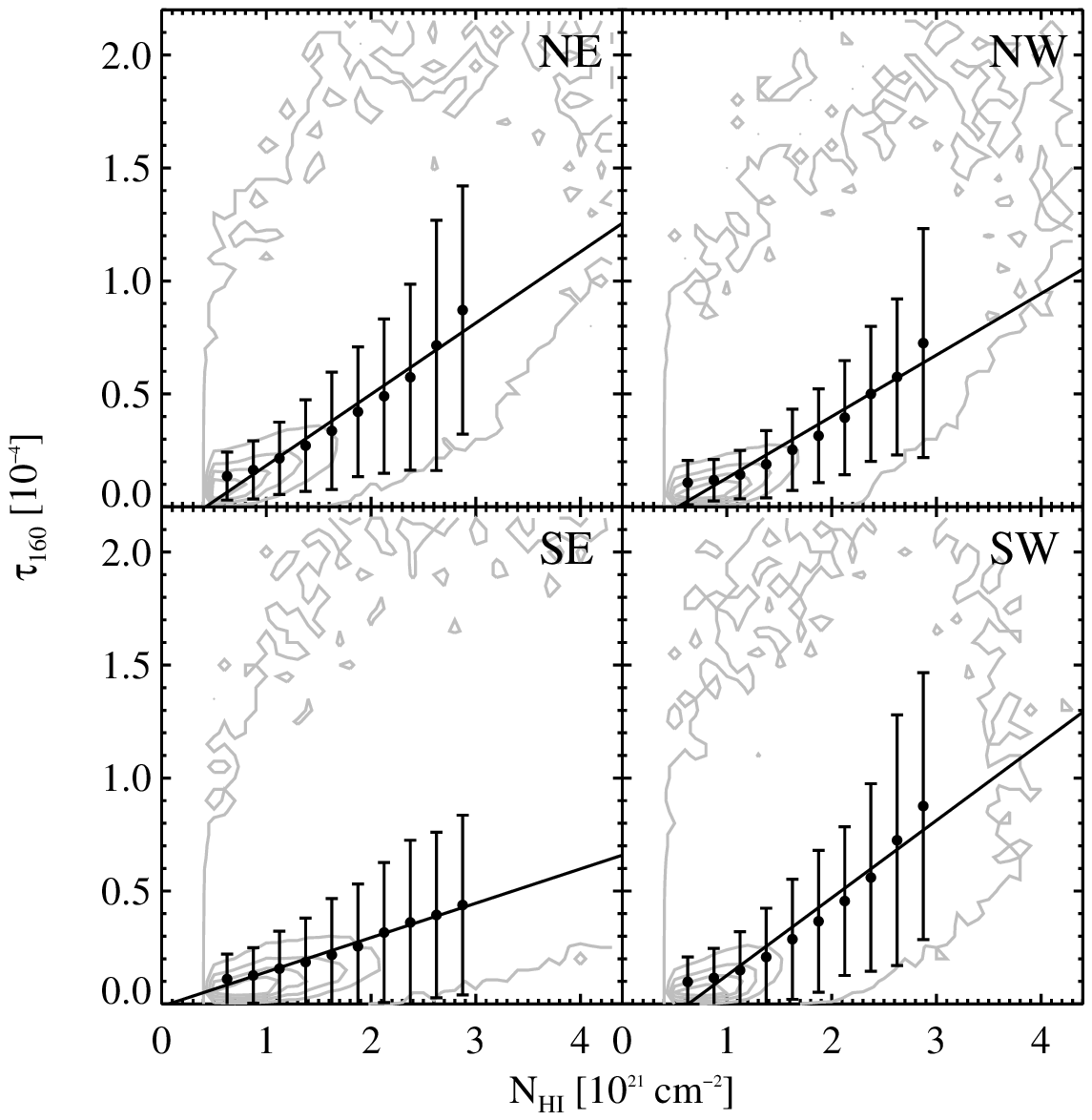}{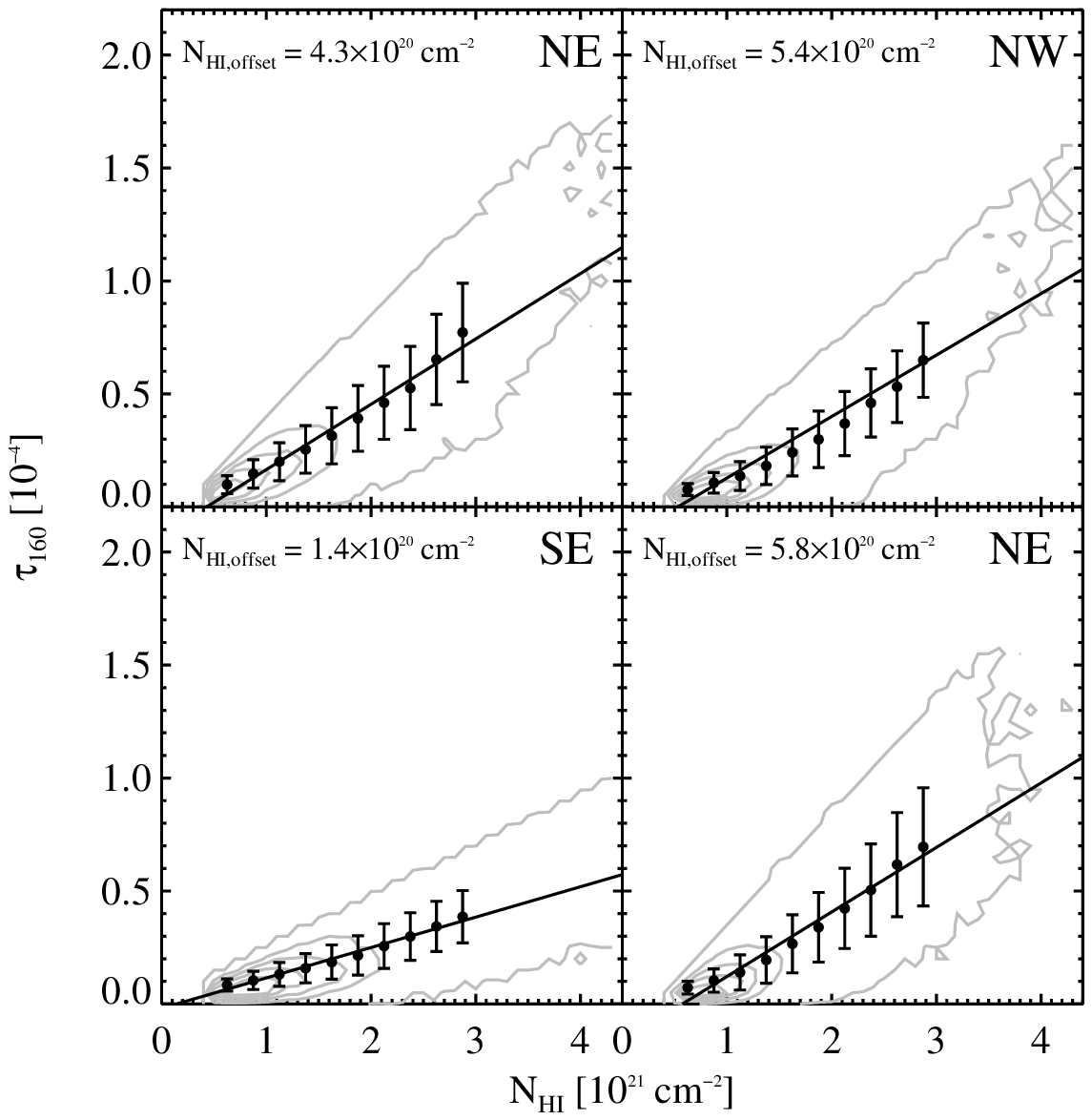}
\caption{Relationships between $\tau_{160}$ and the column density of \hi\ for the quadrants of the LMC using our fiducial $\beta=1.8$ dust modeling results. The contours show the density of points at 80\%, 60\%, 40\%, and 20\% of the maximum with the outer contour showing the full extent of the distribution. The black points show the median $\tau_{160}$ values with $1\sigma$ in $2.5\times10^{20}$ cm$^{-2}$ bins from $1-3\times10^{21}$ cm$^{-2}$ (chosen to avoid low number of values and the threshold where CO-dark \htwo\ can exist). The black lines show the fit to the medians, which are used to make rough estimates of the \htwo\ and to determine the \hi\ offset. The left set of plots have only regions near bright CO masked. The right set of plots have additional regions with significant estimated \htwo\ masked (see Section \ref{subsection:h2_method}) based on the quadrant fits to the binned median values of the distribution shown in the plots on the left.
\label{fig:lmc_Nhi_tau_quadrants_beta}}
\end{figure*}

\begin{figure*}[tp]
\epsscale{1}
\plottwo{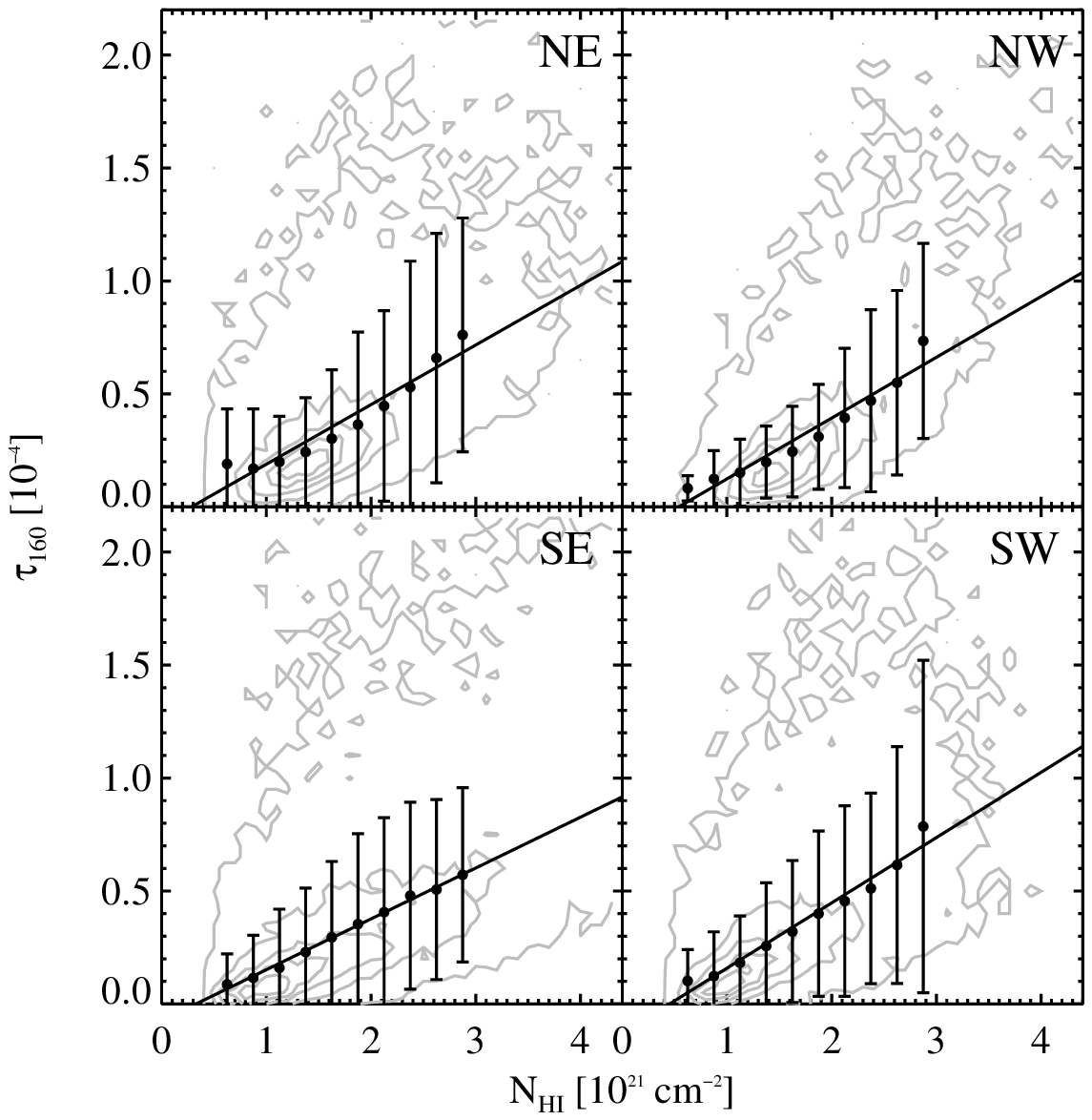}{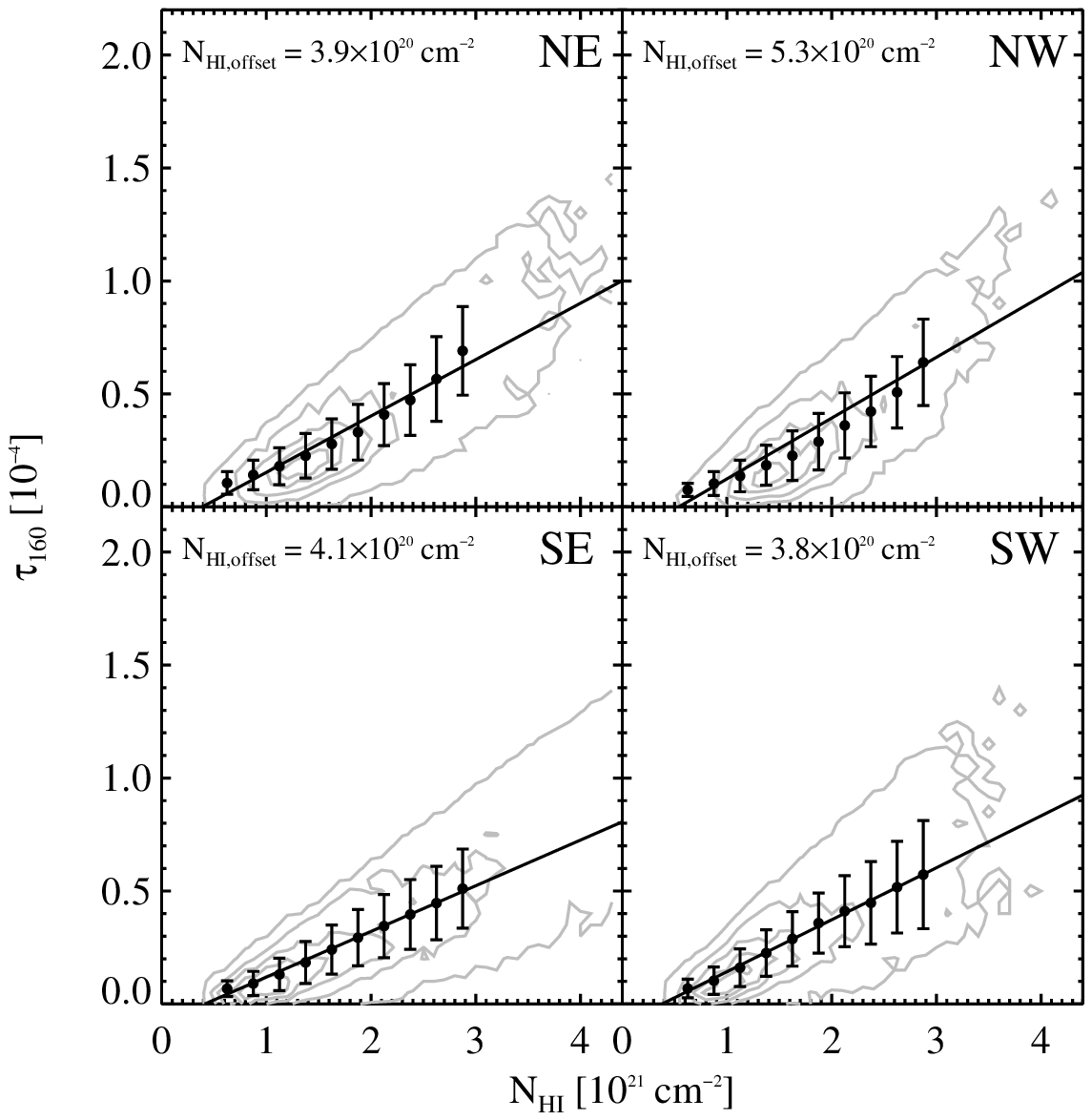}
\caption{Same as Figure \ref{fig:lmc_Nhi_tau_quadrants_beta} but with $\tau_{160}$ from the \citet{gor14} dust modeling results. 
\label{fig:lmc_Nhi_tau_quadrants_bembb}}
\end{figure*}

\begin{figure*}[tp]
\epsscale{1}
\plotone{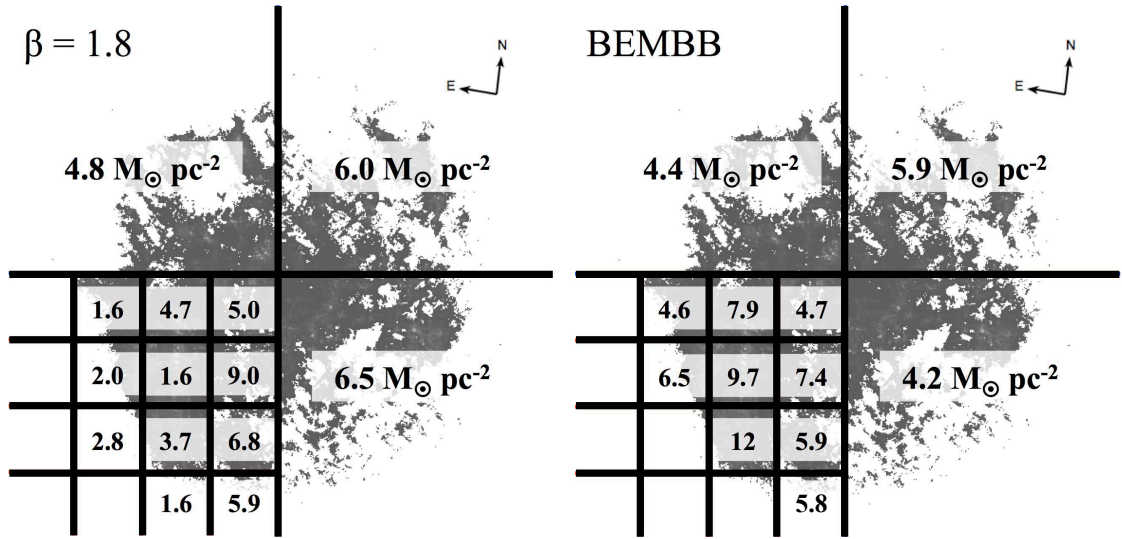}
\caption{The \hi\ offset fit for each region in the LMC for the fiducial $\beta$=1.8 dust modeling in this work ($left$) and \citet{gor14} ($right$).The background grey-scale shows the $tau_{160}$ map from the \citet{gor14} modeling.
\label{fig:lmc_regions}}
\end{figure*}

\section{Diffuse \ha}
\label{appendix:diffuse_ha}

The H$\alpha$ images of the Magellanic Clouds clearly show extended, low-level of emission throughout the galaxies, which traces the ISM component referred to as diffuse ionized gas (DIG) or the warm ionized medium (WIM). In the Milky Way, the filling factor of the WIM ranges from $0.1-0.4$, with evidence that it increases with distance from the mid-plane \citep{ber06}, and contribute $\sim{10-15\%}$ of the total \hii\ emission \citep{rey93}. More detailed studies of the WIM in the Milky Way find that physical conditions differ from conditions in classical \hii\ regions vary widely \citep{haf09}. Early work by \citet{ken86} demonstrated that an extended \ha\ component accounts for $\sim15\%$ of the flux in the LMC. Later, \citet{ken95} found that the diffuse component of the \ha\ emission, found by comparing the total flux to the integrated flux from \hii\ regions, is $\sim25-35\%$ in the LMC and $\sim34-40\%$ in the SMC. These fractions of diffuse or extended emission are consistent with what is found for Magellanic Irregulars \citep{ken89} and in spirals, including the $20-40\%$ fraction found in M31 \citep{wal94}. The diffuse \ha\ component is similar to other star-forming galaxies and a significant fraction of the total emission.

While widespread amongst galaxies, the origin of the diffuse ionized gas is unclear--is it all escaped \ha\ photons from star-forming regions or is the gas ionized within the diffuse ISM (\citealt{rah11} and references therein)? If the gas is primarily ionized within the diffuse ISM by mechanisms not directly related to star formation and not accounted for in the star formation rate calibration, then including the diffuse emission would overestimate the star formation rate. \citet{pel12} studied the optical depth of \hii\ regions in the Magellanic Clouds and found that the luminosity of escaped ionizing radiation provides enough power to ionize the diffuse gas, suggesting that the diffuse \ha\ emission in the Magellanic Clouds could all be escaped radiation from young, massive stars in \hii\ regions. Even if all of the diffuse \ha\ emission can be attributed to star formation, the appropriateness of including the emission in the star formation rate depends on the details of the star formation rate conversion calibration. The calibration by \citet{cal07} was done on scales of $200-600$ pc, which likely includes some extended emission. We include the diffuse \ha\ emission in our analysis and convert the \ha\ maps to star formation rate assuming that massive stars are responsible for all of the \ha\ flux.

\section{Comparison to Previous Dust-Based \htwo\ Estimates}
\label{appendix:prev_work}

\input{table3.tex}

We compare our dust-based molecular gas estimates to similar estimates in the literature and find that all of the estimates are consistent within the uncertainties given differences in methodology and data. Table \ref{table:h2_estimates} provides all of our total molecular gas estimates alongside those from the literature. The early estimates from \citet{isr97} for both the LMC and SMC were based on low resolution $IRAS$ data, did not directly model the dust emission (using instead a scaled far-infrared surface brightness, $\mu_{\textnormal{FIR}}$, based on the difference in dust temperature relative to a fiducial value), and used one effective gas-to-dust ratio for each galaxy based on a few reference positions. These early estimates are likely high due to the lack of long wavelength data, which causes a bias towards higher dust temperature, and therefore high effective gas-to-dust ratios; regardless, the estimates are still with a factor of $\sim{2}$ for the LMC and $\sim{5}$ for the SMC.

\citet{ber08} produced a dust-based estimate of the molecular gas for the LMC using the $Spitzer$ SAGE data and found a total molecular mass of $3.3\times10^{8}$ \msol. They chose a single \dGDR\ equal to the value at the lowest 5\% level of the \dGDR\ distribution, $\tau_{160}/\Nhi=8.8\times10^{-26}$ cm$^{2}$ (or $\Nhi/\tau_{160}=1.1\times10^{25}$ cm$^{2}$). This value of gas-to-gas ratio is consistent with the average values we find in our maps of the gas-to-dust ratio, but our maps have a wide range of values. \citet{ber08} takes this value of the gas-to-dust ratio and applies it to the entire galaxy. The primary difference that drives the higher molecular gas mass estimate is that \citet{ber08} fits lower dust temperatures (median $T_{\text{d}}\sim18$ K) to the $Spitzer$ 160 \micron\ and $IRIS$ 100 \micron\ data. The dust modeling used in this work includes more infrared bands and longer wavelength data and we find a higher average dust temperature of $T_{\text{d}}=23$ K. Changing $T_{\text{d}}$ from 18 K to 23 K (and holding $I_{160}$ constant) results in a decrease in $\tau_{160}$ of a factor of $\sim{3}$. When the similar gas-to-dust ratio is applied to the map of higher values of $\tau_{160}$, the total gas, and therefore molecular gas mass estimate is higher. Scaling our $\tau_{160}$ up by a factor of 3 yields a total molecular gas mass of $2.1\times10^{8}$ \msol, which is comparable to the \citet{ber08} estimate. The \Mmol\ estimate from \citet{ber08} can serve as an upper limit to the amount of gas associated with the excess far-infrared emission.

The most recent estimates for the molecular gas masses come from \citet{rom14}, which also used the dust modeling from \citet{gor14}. While this work focuses on creating maps of the molecular gas, \citet{rom14} studies the global relationship between the gas and dust in the different gas phases and does not explore any spatial variations. The molecular gas mass estimate we show in Table \ref{table:h2_maps} combines the estimate of the molecular gas traced by bright CO emission (using the fiducial $X_{\text{CO}} = 2\times10^{20}$ cm$^{-2}$ (K km s$^{-2}$)$^{-1}$) in the LMC $X_{\text{CO}} = 1\times10^{21}$ cm$^{-2}$ (K km s$^{-2}$)$^{-1}$) in the SMC) with the estimate of the amount of molecular gas not traced by bright CO emission (``CO-dark'' or ``CO-faint''). They estimate the amount of molecular gas not traced by CO by applying the diffuse gas-to-dust ratio (GDR$^{\text{dif}}=380$ in the LMC and GDR$^{\text{dif}}=1200$) to the regions where molecular gas is expected, based on the dust surface density, but no CO is detected. The total molecular gas mass estimates for the LMC and SMC are significantly lower than the estimates from this work. \citet{rom14} uses a constant gas-to-dust ratio while we use a map of the gas-to-dust ratio, the average gas-to-dust ratios from our maps are $\sim{50\%}$ larger than the average values by \citet{rom14}, and they use CO to estimate part of the molecular gas. The main factor driving the lower molecular gas is the low gas-to-dust ratio applied uniformly across the galaxies. The difference in gas-to-dust ratio is largely due to the difficulty in fitting a linear relation to a noisy distribution ($\siggas$-$\sigdust$); \citet{rom14} finds a range in the fitted global GDR$^{\text{dif}}$ of $380-540$ depending on the fitting method. 

The new estimate for the SMC using the $Herschel$ data is lower than the estimates based on $Spitzer$ data: $\sim40\%$ lower than the estimate from \citet{bol11}, and $\sim60\%$ from \citet{ler07}. This is well within the factor of $2-3$ estimated systematic uncertainty from \citet{bol11}. Given the differences in methodology used in all of the previous estimates and their respective levels of uncertainty, we find all of the molecular gas mass estimates to be consistent.

\section{Implementation of KMT+}
\label{appendix:KMT+}
Here we explain how the KMT+ code was implemented with a brief description of the equations involved in the model. The KMT+ model takes the total surface density of gas (\siggas), the volume mass density of the stars and dark matter in the disk ($\rho_{\textnormal{sd}}$), the metallicity normalized to solar ($Z'$), and a ``clumping parameter'' ($f_{c}$ in KMT+, $c$ in KMT09) as input parameters and solves for volume number density of the cold neutral medium ($n_{\textnormal{CNM}}$), the fraction of \htwo\ ($f_{\textnormal{H}_{2}}$), the strength of the radiation field normalized to the solar value ($G'_{0}$), and the star formation rate (\sigdotstar). The model is based on the following equations (equations 8, 9, 10, 14, and 15 from \citet{kru13}). The fraction of \htwo, $f_{\textnormal{H}_{2}}\equiv \sightwo/(\sighi + \sightwo)$, is determined by
\begin{equation}
f_{\textnormal{H}_{2}} = \left\{ 
  \begin{array}{l l}
    1-(3/4)s/(1+0.25s), & \quad s<2 \\
    0, & \quad s \ge{2}
  \end{array} \right. 
\end{equation}
\begin{equation}
s \approx \frac{\ln (1+0.6\chi+0.01\chi^{2})}{0.6\tau_{c}},~\tau_{c} = 0.066f_{c}Z'\Sigma_{0}
\end{equation}
\begin{equation}
\chi =  7.2\frac{G'_{0}}{n_{1}},~n_{1} = \frac{n_{\textnormal{CNM}}}{10\textnormal{ cm}^{-3}}.
\end{equation}
With a value for $f_{\textnormal{H}_{2}}$, we can then calculate the star formation rate per unit area,
\begin{equation}
\sigdotstar = f_{\textnormal{H}_{2}}\epsilon_{\textnormal{ff}}\frac{\Sigma}{t_{\textnormal{ff}}}~\textnormal{with}~
t_{\textnormal{ff}} \approx \frac{\pi^{1/4}}{\sqrt{8}}\frac{\sigma_{g}}{G(\Sigma^{3}_{\textnormal{GMC}}\Sigma)^{1/4}},
\end{equation}
where $\epsilon_{\textnormal{ff}}\approx0.01$, $t_{\textnormal{ff}}$ is the free-fall time of the molecular gas, $\sigma_{g}\approx8~\kms$ is the velocity dispersion of the galactic disc and $\Sigma_{\textnormal{GMC}}\approx 85 \msol$ pc$^{-2}$ is the characteristic surface density of self-gravitating molecular clouds.
At the same time, the radiation field is proportional to the surface density of the star formation rate (following OML10), 
\begin{equation} 
G'_{0} \approx \frac{\sigdotstar}{\dot{\Sigma}_{*,0}}
\end{equation}
with the normalization set by the conditions in the solar neighborhood, $\dot{\Sigma}_{*,0}=2.5\times10^{-3}$ \msol pc$^{-2}$ Myr$^{-1}$. 

The main ansatz of the KMT+ model is that volume density of the cold neutral medium required by hydrostatic equilibrium represents a floor to the possible density, hence the need to calculate both the density from two-phase equilibrium ($n_{\textnormal{CNM,2p}}$) and the density from hydrostatic equilibrium ($n_{\textnormal{CNM,hydro}}$) and take the maximum:
\begin{equation}
n_{\textnormal{CNM}} = \mbox{max}(n_{\textnormal{CNM,2p}},n_{\textnormal{CNM,hydro}}).
\end{equation}
The density of the CNM in two-phase equilibrium comes from the KMT09 model and depends on the radiation field and the metallicity,
\begin{equation}
n_{\textnormal{CNM,2p}} \approx 23 G'_{0} \left( \frac{1+3.1Z'^{0.365}}{4.1} \right)^{-1}\textnormal{ cm}^{-2}.
\end{equation}
The minimum CNM density from hydrostatic equilibrium requires the thermal pressure ($P_{\textnormal{th}}$) and the maximum temperature of the CNM ($T_{\textnormal{CNM,max}}$),
\begin{equation}
n_{\textnormal{CNM,hydro}} = \frac{P_{\textnormal{th}}}{1.1k_{B}T_{\textnormal{CNM,max}}}.
\end{equation}
The maximum temperature the CNM can have and still exist is taken to be $T_{\textnormal{CNM,max}}\approx243$ K, from the simple analytic model by \citet{wol03}. The thermal pressure equation follows OML10 and calculates the pressure contributions from the gravity contributions from the gas phases and the stellar and dark matter density, 
\begingroup
  \thinmuskip=1mu
  \medmuskip=2mu minus 2mu
  \thickmuskip=3mu
  \begin{equation}
  	\scalebox{.8}
	{$\displaystyle P_{\textnormal{th}} = \frac{\pi G \sighi^{2}}{4\alpha} \left\{ 1+2R_{\textnormal{H}_{2}}+ \left[(1+ 	2R_{\textnormal{H}_{2}})^{2}+\frac{32\zeta_{d}\alpha\tilde{f_{w}}c_{w}^{2}\rho_{\textnormal{sd}}}{\pi G 		\sighi^{2}} \right]^{1/2}\right\}$}
   \end{equation}
\endgroup
where $R_{\textnormal{H}_{2}}\equiv\sightwo/\sighi=f_{\textnormal{H}_{2}}/(1-f_{\textnormal{H}_{2}})$, $\zeta_{d}\approx0.33$ is a numerical factor whose exact value depends on the shape of the gas surface density profile, $\rho_{\textnormal{sd}}$ is the volume density of stars and and dark matter, $c_{w}\approx8~\kms$ is the sound speed in the warm neutral medium, $\tilde{f_{w}}=0.5$ is the adopted fiducial value for the ratio of the mass-weighted mean square thermal velocity dispersion to the square of the warm gas sound speed, $\alpha\approx5$ is the ratio of total pressure in the mid-plane to the thermal pressure ($P_{\textnormal{th}}$) due to the additional support provided by turbulence, magnetic fields, and cosmic ray pressure.

To solve these sets of equations, we first guess a value of $G'_{0}$ and then a value of $f_{\textnormal{H}_{2}}$ and iterate through both until, first, a self-consistent value of $f_{\textnormal{H}_{2}}$ is found (inputing the guess into (14) and checking against the value from (5)). We then evaluate \sigdotstar using (8) and check our original guess for $G'_{0}$ against (9) and iterate until we have self-consistent values for $G'_{0}$ and $f_{\textnormal{H}_{2}}$ (within some threshold). For the following predictions using the KMT+ model, we input a map of $\rho_{\textnormal{sd}}$ based on the dark matter profile from rotation curves and stellar density from the $Spitzer$ 3.6 \micron\ maps using the conversion from \citet{ler08}.

\end{document}

